\documentclass[a4paper,12pt]{spieman}  
\usepackage[pdftex]{graphicx} 
\usepackage{setspace}
\usepackage{tocloft}

\title{On-ground calibration of the X-ray, gamma-ray, and relativistic electron detector onboard TARANIS}

\author[a*]{Yuuki~Wada}
\author[b,c]{Philippe~Laurent}
\author[c]{Damien~Pailot}
\author[b]{Ion~Cojocari}
\author[c]{Eric~Br\'{e}elle}
\author[c]{St\'{e}phane~Colonges}
\author[c]{Jean-Pierre Baronick}
\author[b,c]{Fran\c{c}ois~Lebrun}
\author[d]{Pierre-Louis~Blelly}
\author[e]{David~Sarria}
\author[f]{Kazuhiro~Nakazawa}
\author[c]{Miles~Lindsey-Clark}

\affil[a]{Division of Electrical, Electronic, and Infocommunications Engineering, Graduate School of Engineering, Osaka~University, 2-1 Yamadaoka, Suita, 565-0871 Osaka, Japan}
\affil[b]{Service d'Astrophysique, IRFU/DRF, CEA Saclay, 91191 Gif-sur-Yvette Cedex, France}
\affil[c]{Laboratoire Astroparticule et Cosmologie, Universit\'{e} de Paris/CNRS, 10 rue Alice Domon et L\'{e}onie Duquet, 75205 Paris Cedex 13, France}
\affil[d]{Institut de Recherche d'Astrophysique et Plan\'{e}tologie, 9 avenue du Colonel Roche, BP 44346, 31028 Toulouse Cedex 4, France}
\affil[e]{Birkeland Centre for Space Science, University of Bergen, Postboks 7803, N-5020 Bergen, Norway}
\affil[f]{Kobayashi-Maskawa Institute for the Origin of Particles and the Universe, Nagoya University, Furo-cho, Chikusa-ku, Nagoya, 464-8601 Aichi, Japan}

\cftpagenumbersoff{figure}
\cftpagenumbersoff{table} 

\begin{document} 
\maketitle

\begin{abstract}
	We developed the X-ray, Gamma-ray and Relativistic Electron detector (XGRE) onboard the TARANIS satellite, 
	to investigate high-energy phenomena associated with lightning discharges 
	such as terrestrial gamma-ray flashes and terrestrial electron beams.
	XGRE consisted of three sensors. Each sensor has one layer of LaBr$_{3}$ crystals for X-ray/gamma-ray detections,
	and two layers of plastic scintillators for electron and charged-particle discrimination.
	Since 2018, the flight model of XGRE was developed, and validation and calibration tests, such as a thermal cycle test 
	and a calibration test with the sensors onboard the satellite were performed before the launch of TARANIS on 17 November 2020.
	The energy range of the LaBr$_{3}$ crystals sensitive to X-rays and gamma rays was determined 
	to be 0.04--11.6~MeV, 0.08--11.0~MeV, and 0.08--11.3~MeV for XGRE1, 2, and 3, respectively.
	The energy resolution at 0.662~MeV (full width at half maximum) was to be 20.5\%, 25.9\%, and 28.6\%, respectively.
	Results from the calibration test were then used to validate a simulation model of XGRE and TARANIS.
	By performing Monte Carlo simulations with the verified model, 
	we calculated effective areas of XGRE to X-rays, gamma rays, electrons, 
	and detector responses to incident photons and electrons coming from various elevation and azimuth angles.

\end{abstract}


{\noindent \footnotesize\textbf{*}Yuuki Wada, \linkable{wada@yuuki-wd.space}}

\begin{spacing}{1}

\section{Introduction}\label{sect:intro}  

	While lightning discharges are well-known atmospheric phenomena, recent studies have revealed that they are also associated with
	transient luminous events (TLEs\cite{Pasko_2012}) and high-energy phenomena\cite{Dwyer_2012b}.
	Terrestrial gamma-ray flashes (TGFs) are high-energy atmospheric phenomena coincident with lightning discharges\cite{Fishman_1994,Smith_2005}.
	They have typical durations ranging from tens to hundreds of microseconds \cite{Foley_2014}, and the energy of gamma rays extends up to $>$20~MeV\cite{Smith_2005}.
	Since the discovery\cite{Fishman_1994} made by the Burst and Transient Source Experiment (BATSE) onboard the Compton Gamma-Ray Observatory (CGRO),
	TGFs have been detected by in-orbit satellites such as the Reuven Ramaty High-Energy Solar Spectroscopic Imager\cite{Smith_2005} (RHESSI), 
	the Gamma-ray Burst Monitor (GBM) of Fermi\cite{Briggs_2010,Mailyan_2016}, Astro-Rivelatore Gamma a Immagini Leggero\cite{Marisaldi_2010,Tavani_2011} (AGILE), BeppoSAX\cite{Ursi_2017}, 
	and the Atmospheric-Space Interaction Monitor\cite{Neubert_2019b,Ostgaard_2019b} (ASIM) onboard the International Space Station.

	TGFs are thought to be bremsstrahlung photons from relativistic electrons, accelerated by strong electric fields in lightning.
	The relativistic runaway electron avalanche\cite{Gurevich_1992} (RREA) is a promising theory for electron acceleration and multiplication by electric fields in the dense atmosphere.
	On the other hand, gamma-ray fluxes measured by satellite observations cannot be explained by electron-multiplication factors achieved by the RREA process\cite{Dwyer_2008}.
	While more efficient electron-multiplication mechanisms such as the cold-breakdown \cite{Carlson_2010b,Celestin_2011} and the relativistic feedback models\cite{Dwyer_2003b,Dwyer_2012a} have been proposed, 
	definitive solutions have not been reached yet.

	Among TGF candidates detected by BATSE/CGRO, RHESSI, AGILE, Fermi-GBM, and ASIM, some events have a duration longer than 1~ms.
	They are now interpreted as terrestrial electron beams (TEBs), rather than TGFs\cite{Dwyer_2008b,Briggs_2011,Sarria_2021}.
	Gamma-ray photons of a TGF can create electrons by Compton scatterings and pair productions, and positrons by pair productions in the atmosphere.
	The produced electrons and positrons sometimes reach the satellites and are detected as a TEB. 
	The electrons and positrons can be trapped by and move along a magnetic-force line of the earth. 
	Since electrons travel various distances along the magnetic fields, the duration of TEBs is typically longer than that of TGFs.
	In some cases, TEBs can be detected far from the parent lightning discharges.
	TEBs with enough positrons make a significant peak of the annihilation line at 511~keV in their energy spectrum\cite{Briggs_2011}.

	One of the powerful tools to reveal the source mechanism of TGFs is multi-wavelength observations.
	Radio-frequency measurements are useful to observe parent lightning discharges of TGFs, 
	and have revealed that TGFs were produced during leader progression\cite{cummer_2015}
	and/or coincident with distinct classes of low-frequency pulses called energetic in-cloud pulses\cite{Lyu_2015,Lyu_2018} (EIPs) and slow pulses\cite{Pu_2019}.
	Also, ASIM has optical cameras/photometers and X/gamma-ray detectors, and investigates the relation between TGFs and lightning flashes\cite{Lindanger_2022}, 
	and between TGFs and TLEs\cite{Neubert_2019b,Bjorge-Engeland_2022}.

	The Tool for the Analysis of RAdiation from lightNIngs and Sprites (TARANIS) was a satellite mission developed by the National Centre for Space Studies of France (CNES),
	to investigate the relation between lightning discharges and TLEs, TGFs, and TEBs\cite{Lefeuvre_2008}. TARANIS was a 200-kg satellite with the Myriade platform of the CNES, 
	and consisted of six instruments, MicroCameras and Photometers (MCP), X-ray, Gamma-ray and Relativistic Electron detector (XGRE), 
	Instrument D\'{e}tecteurs d'Electrons Energ\'{e}tiques (IDEE; energetic electron detector), 
	Instrument de Mesure du champ Electrique-Basse Fr\'{e}quence (IME-BF; an instrument for low-frequency measurements), 
	Instrument de Mesure du champ Electrique-Haute Fr\'{e}quence (IME-HF; an instrument for high-frequency measurements),
	and Instrument de Mesure du champ Magn\'{e}tique (IMM; an instrument for magnetic wave measurements), with the Multi Experiment Interface Controller equipment (MEXIC).
	The satellite was developed in collaboration with the Laboratory of Physics and Chemistry for Environment and Space (LPC2E),
	Alternative Energies and Atomic Energy Commission (CEA), Research Institute in Astrophysics and Planetology (IRAP),
	Laboratory for Atmospheres and Space Observations (LATMOS), Laboratory Astroparticles and Cosmology (APC), 
	Stanford University in the US, Institute of Atmospheric Physics and Charles University in the Czech Republic and Space Research Centre of Polish Academy of Sciences.

	At 22:52, 17 November 2020 in Universal Coordinated Time (UTC), TARANIS was launched by the VEGA17 rocket of Ariane Space from the Kourou Space Center in French Guiana.
	TARANIS was planned to be transported into a sun-synchronous orbit at a 700-km altitude. 
	However, the launch failed due to a malfunction at the 4th stage of the launcher, and unfortunately, the mission was lost.
	While the scientific mission that TARANIS was aiming at cannot be achieved, the development process of TARANIS is of great importance for future space missions.
	In the present paper, we report the on-ground validation and calibration campaign of the XGRE detector, 
	and its performance verification with Monte Carlo simulations. Throughout the paper, the time is described in UTC.

\section{XGRE Detector}\label{sect:det}

	XGRE was the core detector on TARANIS for studying high-energy phenomena in lightning.
	It was designed to be sensitive to X-rays, gamma rays, and relativistic electrons to detect mainly TGFs and TEBs, which are mainly caused by lightning discharges.	
	It was also sensitive not only to high-energy radiation from lightning discharges but also to one from space such as gamma-ray bursts (GRBs).
	The mission-required performance was to have sensitivity to X-rays and gamma rays of 0.02--10~MeV and electrons of 1--10~MeV.
	Since X-rays below 100~keV are susceptible to attenuation in the atmosphere, 
	their energy spectra can be used to estimate the generation height of TGFs. The XGRE detector was mainly developed by APC/Paris.

	\begin{figure}
		\begin{center}
		\begin{tabular}{c}
		\includegraphics[width=\linewidth]{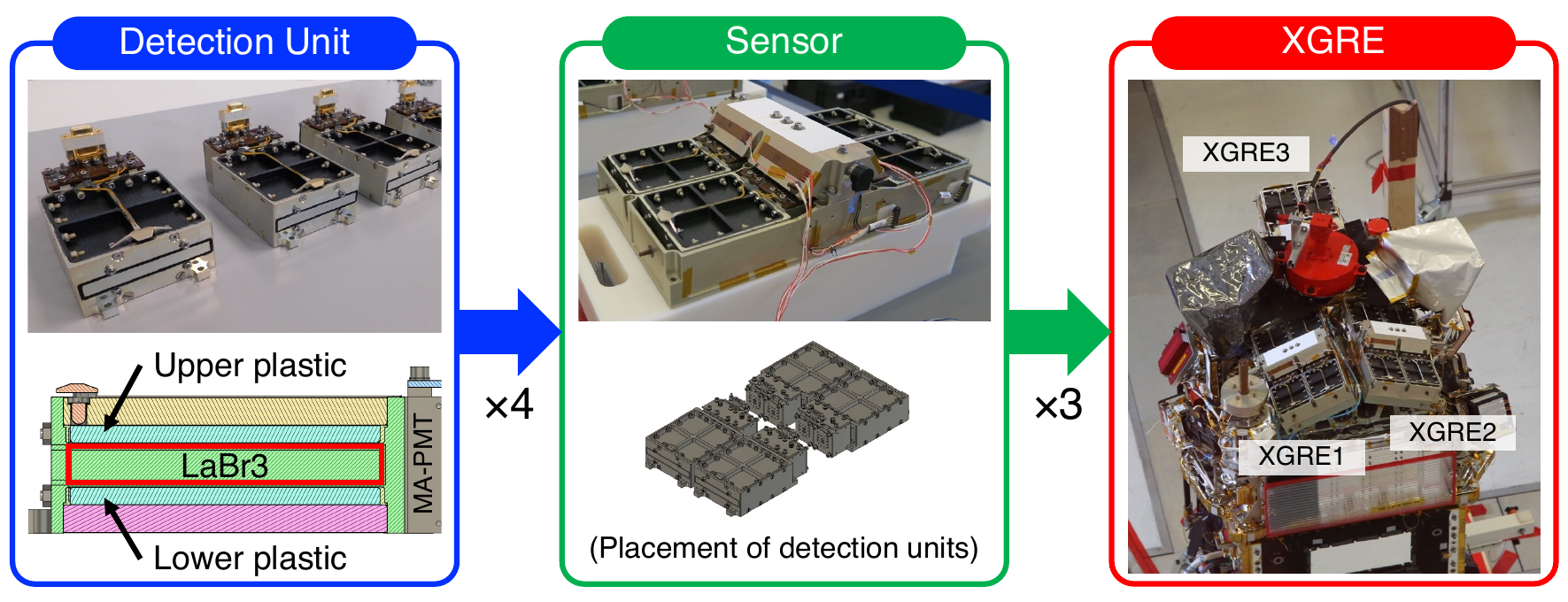}
		\end{tabular}
		\caption{The composition of the XGRE detector.\label{fig:structure}}
		\end{center}
	\end{figure} 

	\begin{figure}
		\begin{center}
		\begin{tabular}{c}
		\includegraphics[width=\linewidth]{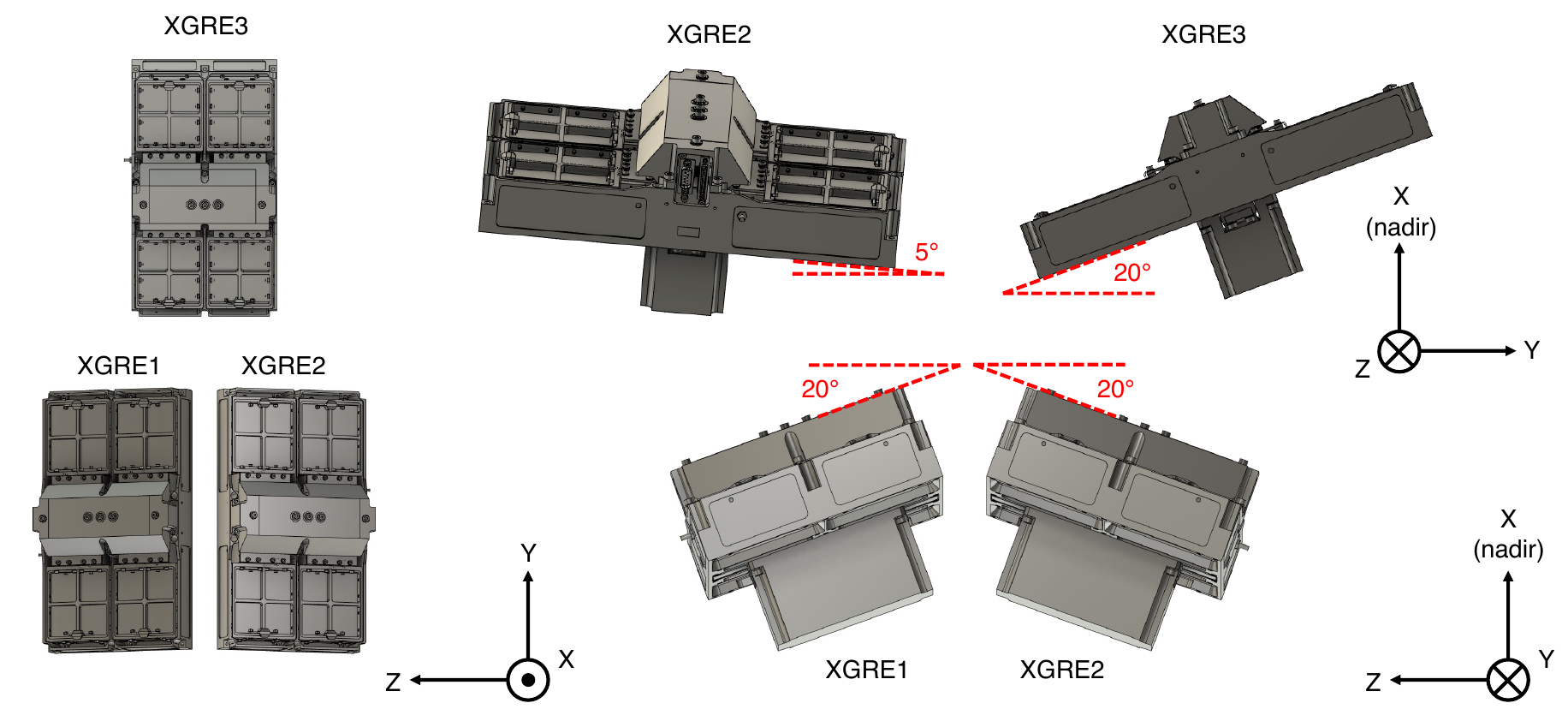}
		\end{tabular}
		\caption{The mounting geometry of the XGRE detector.\label{fig:dimension}}
		\end{center}
	\end{figure} 

	XGRE consisted of three sensors. Figure~\ref{fig:structure} shows the configuration and composition of XGRE, and the three sensors mounted on the satellite. 
	The sensors were mounted in the nadir direction of the satellite (Earthside), and were most sensitive to X-rays, gamma rays, and electrons emitted directly upward from the Earth. 
	Each sensor was identical, but the angle at which it was mounted was different. Figure~\ref{fig:dimension} shows the mounting angle of the sensors. 
	XGRE1 and XGRE2 were tilted 20$^{\circ}$ around the Y~axis and 5$^{\circ}$ around the Z~axis. XGRE3 was tilted 20$^{\circ}$ around the Z~axis. 
	Since the effective area of XGRE changes with respect to the incoming direction of X-rays and gamma rays due to the difference in the mounting angle of each sensor, 
	the difference in count rates of the sensors can be used to estimate the source location of TGFs.

	Each sensor consisted of four detection units. Each detection unit contained an 84$\times$90$\times$8.7~mm$^{3}$ LaBr$_{3}$ scintillation crystal (manufactured by Saint-Gobain). 
	Due to its hygroscopic nature, the LaBr$_{3}$ crystal was housed in aluminum with a thickness of 2~mm. 
	Two 84$\times$90$\times$5~mm$^{3}$ plastic scintillators (BC400 by Saint-Gobain) were arranged to sandwich the LaBr$_{3}$ crystal. 
	Figure~\ref{fig:structure} shows the appearance and cross-section of detection units. 
	The nadir side was the upper plastic, and the space side was the lower plastic scintillator. 
	Particle discrimination can be performed by taking coincidence counting of the LaBr$_{3}$ crystal and the plastic scintillators; 
	X-rays or gamma rays when each crystal is hit alone; electrons within the measurable energy range when the LaBr$_{3}$ and one plastic scintillators are hit simultaneously;
	and high-energy electrons, positrons, and other charged particles exceeding the measurable energy range when three crystals are hit simultaneously.

	Scintillation photons from each scintillator were read out from the side with two Hamamatsu multi-anode photomultipliers (MA-PMTs) R8900. 	
	R8900 is a 16-pixel MA-PMT, with 4 channels in the upper row for reading out the upper plastic scintillator, 
	8 channels in the middle two rows for LaBr$_{3}$, and the lower row for reading out the lower plastic scintillator. 
	Each XGRE sensor has two electronic boards, the Analog Front-End (FEA) and the Digital Front-End (FEN). 
	The FEA has three separate analog chains corresponding to the upper/lower plastics and the LaBr$_{3}$ scintillator readout. 
	The signals coming from the 4 detection units forming one sensor are connected together, 
	thus the dead-time percentage is dictated by the particle flux incident on the sensor. 
	A stand-alone charge preamplifier A250F (Amptek) and a CR-RC shaper form each analog chain. 
	The preamplifier is equipped with an active reset, which set a fixed deadtime of 350 ~ns, 
	and the gain is tailored to the light output of the different scintillators. 
	On the FEN, dedicated radiation-hard ADCs are used, with a 14~bits resolution for the LaBr$_{3}$ read-out and 12~bits for the plastics. 
	The read-out sequence is controlled by an FPGA that manages the pre-amplifier reset, the ADC operation and the communication with the dedicated XGRE Analyzer (XGA) in MEXIC. 
	The final data packets are generated by XGA. XGA extracted and registered the peak value, detection time, sensor type, and scintillator type for pulse signals above a threshold.
	During normal operation, only the energy spectrum and light curves of each sensor and each scintillator was saved at regular intervals.
	Once XGA detected a sudden increase in count rate due to high-energy phenomena such as TGFs, TEBs, or GRBs, 
	the peak value, detection time, sensor type, and scintillator type were saved and sent to the onboard computer of the satellite. 
	Besides, one sensor had two high voltage (HV) modules. One HV module supplies high voltage to two detection units, i.e. four MA-PMTs.

\section{On-ground Validation and Calibration}\label{sect:cal}

	The on-ground validation and calibration tests were performed in three steps: with the detection units, with the sensors, and with the satellite. 
	A total of 16 flight models of the detection units were produced, and validation tests were conducted at APC in February 2018. 
	In the test, we measured the variation in gain due to the characteristics of MA-PMTs, the analog circuit, and the optical bonding. 
	Then we selected 12 of the 16 units with better performance as flight models and 4 units as spare models. 

	After the sensors were validated at APC, a thermal-cycle test using the flight models was conducted from April to May 2018,
	and then a thermal-vacuum test using the spare sensor was conducted from June to July 2018. 
	These tests were performed at the Laboratory of Space Studies and Instrumentation in Astrophysics (LESIA) of the Paris Observatory.
	In normal cases of satellite development, a thermal-vacuum test is performed with a flight model to validate functions in the space environment.
	On the other hand, we did not perform the thermal-vacuum test on the flight models as the LaBr$_{3}$ crystals used in the detection units is a highly hygroscopic inorganic crystal, and hence an aluminum housing is indispensable to prevent this. 
	However, we discovered that the vacuum test may reduce the air tightness of the housing, and humid air may enter after the vacuum test, degrading the detector's performances 
	(of course, even if the housing tightness is reduced in space, this cannot occur due to the absence of air). 
	Therefore, the flight models were subjected only to a thermal cycle test in a chamber filled with dry nitrogen gas. As an alternative, a thermal-vacuum test was performed on the spare sensor, test which revealed to be finally successful. 

	In November 2018, the validation tests of XGRE alone were completed, and the XGRE detector was handed over from APC to CNES in Toulouse. 
	After the XGRE installation on the satellite (March-April 2019) and vibration and acoustic tests on the entire satellite (June 2019),
	a calibration test of XGRE with the entire satellite was performed using radiation sources.
	Although the launch was delayed due to the COVID-19 lockdown and the failure of VEGA15, 
	the satellite was transported from Toulouse to the Guiana Space Center in September 2020. 	

	A thermal vacuum test of the entire satellite was conducted from January to February 2019, 
	but the XGRE flight model was not mounted on the satellite at that time, instead, a dummy load was mounted.
	In the present paper, we report in detail the temporal resolution verification and the thermal-cycle test for the flight-model sensors, and the calibration test with the sensors onboard the satellite.

\subsection{Verification of Temporal Resolution with Detection Units}\label{sect:timing}

	TGFs are a phenomenon in which photons arrive almost instantaneously (from tens to hundreds of microseconds). 
	It causes pile-ups, which are superpositions of photon pulses.
	When a pile-up occurs, the measurement of the number and energy of photons becomes inaccurate, and it should be avoided as much as possible. 
	Therefore, by dividing one sensor into four detection units, XGRE is designed to reduce the number of photons per a detection unit and reduce pile-ups.
	We verified the photon measurement ability of the detection unit alone in a high count-rate environment.

	A single detection unit was used as the experimental setup. A scintillator was not attached to the PMT, and pulsed light using an LED was used instead of scintillation light.
	Of the pixels of the MA-PMT, the upper and lower rows that read out the plastic scintillators were masked, and the LED light was set to be irradiated on the two middle rows that read out LaBr$_{3}$.
	LED pulses were emitted for exactly 1 second at frequencies of 1~kHz, 10~kHz, 100~kHz, 300~kHz, 500~kHz, 1~MHz, 1.5~MHz, and 2~MHz. 
	The pulses are not generated at random intervals but at regular intervals. Three types of LED brightness were tested, corresponding to 670~keV, 5.0~MeV, and 9.7~MeV equivalent.
	The duration of LED pulses was 75~ns, enough short compared to the shaping time of the preamplifier and considered as a delta function. 

	Regardless of the brightness of the LED, it is confirmed to record pulses up to 1 MHz without missing any pulses.
	Also, when using LED pulses equivalent to 670 keV and 5.0 MeV, it was possible to obtain pulses up to 1.5 MHz
	(testing at 1.5 MHz at 9.7 MeV may cause overcurrent in the high-voltage module, so not tested).
	In the case of LED pulses equivalent to 670 keV, measurements were also performed at 2 MHz, but pile-up occurred and only about 80\% of the pulses could be counted.
	Therefore, the pulse measurement capability was demonstrated up to 1.5 MHz.

	Excessive current may flow between the dynodes of a photomultiplier at high count rates, which can cause a drop of high voltage due to current limits of the high-voltage supply module.
	It causes gain variation, which may affect the observation results. 
	In addition, large current discharges capacitors in PMT dividers excessively, which also causes significant gain changes.
	Therefore, by changing the brightness and frequency of the LED pulses with the same setup above,
	we investigated the gain variation, namely variation in the peak value of the amplifier outputs.

\begin{figure}
		\begin{center}
		\begin{tabular}{c}
		\includegraphics[width=0.7\linewidth]{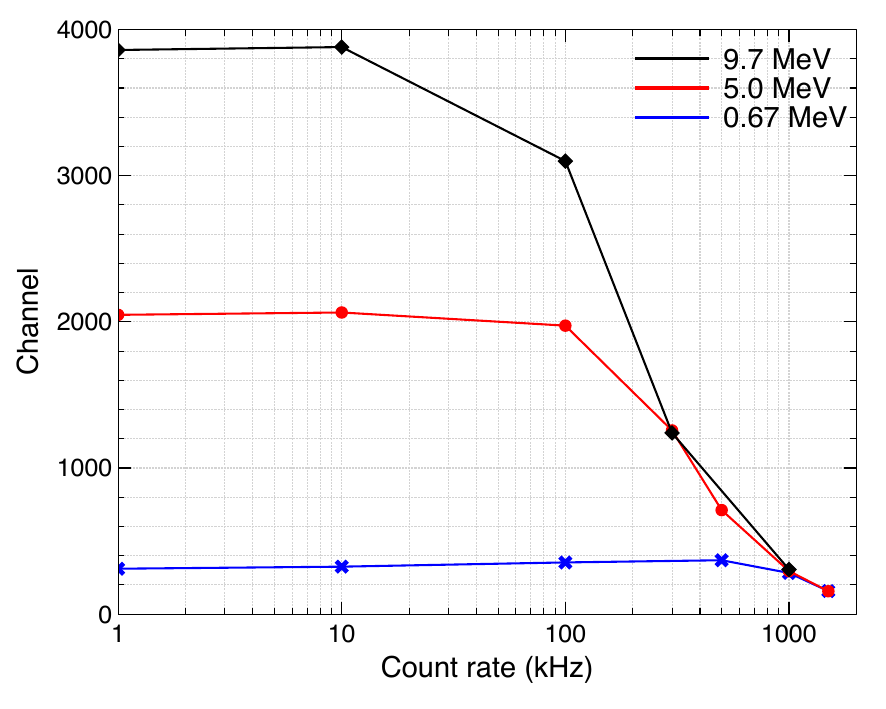}
		\end{tabular}
		\caption[The relation between pulse frequency and peak height.]
		{The relation between pulse frequency and peak height.\label{fig:gain_change}}
		\end{center}
	\end{figure}

	The results are shown in Figure~\ref{fig:gain_change}. With all the brightnesses of LED, a decrease in gain was observed at count rates of 1~MHz or higher.
	In the case of 0.67 MeV equivalent, no gain fluctuation was observed up to 500~kHz, but the gain decreased slightly at 1~MHz.
	With 5.0-MeV equivalent pulses, the gain was constant up to 100~kHz, but a decrease was seen above 300~kHz.
	This is more noticeable in the case of 9.7~MeV, where the gain was constant up to 10~kHz, but decreases above 100~kHz.
	Note that this situation assumes a uniform train of pulses. On the other hand, the arrival time of photons from TGFs is usually random,
	and the maximum count rate in actual observation situations could be degraded (by approximately one third to half).

\subsection{Thermal Cycle Test of Sensors}\label{sect:tct}

	\begin{figure}
		\begin{center}
		\begin{tabular}{c}
		\includegraphics[width=0.7\linewidth]{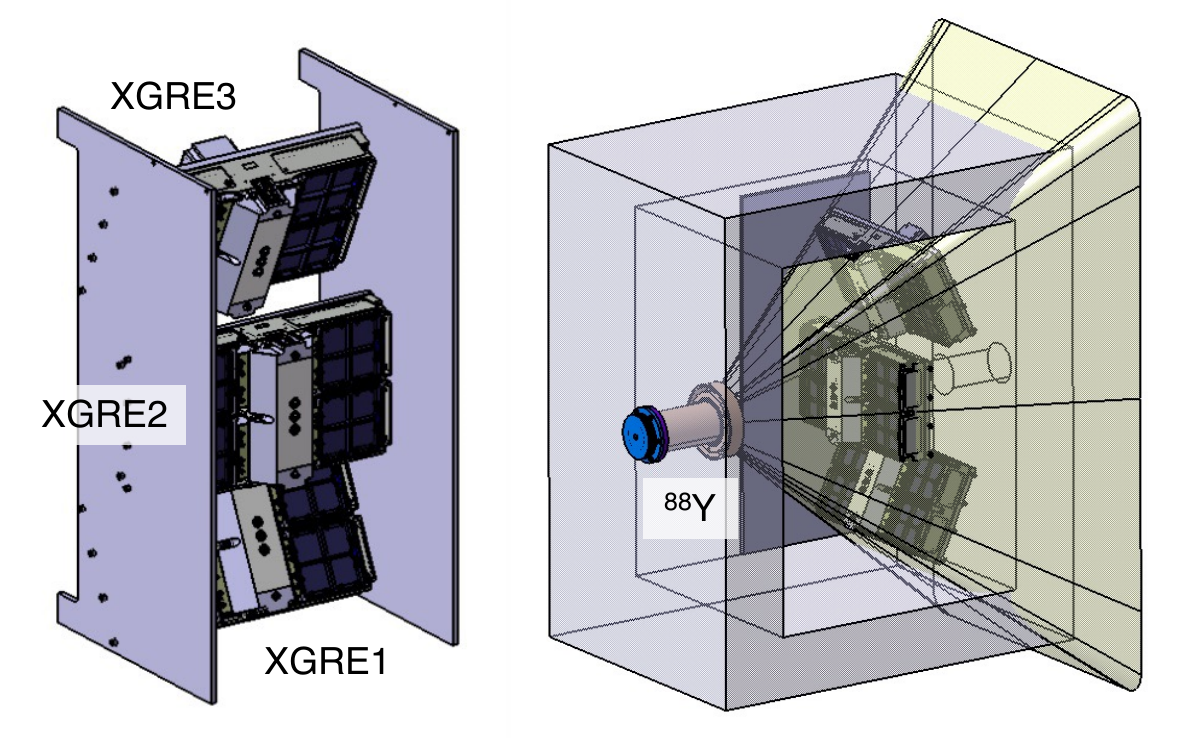}
		\end{tabular}
		\caption[The placement of three sensors during the thermal cycle test.]
		{The placement of three sensors during the thermal cycle test. Left: The sensors with a dedicated mounting jig. Right: Geometry of the radiation source.\label{fig:cycle_geo}}
		\end{center}
	\end{figure} 

	The thermal-cycle test was conducted to verify the operation of XGRE in thermal conditions of the space environment, 
	and to investigate the characteristics of gain variation due to temperature changes. 
	The test was performed with a thermal-cycle chamber EXCAL 2221-T/H at LESIA.
	As shown in the left panel of Figure \ref{fig:cycle_geo}, three sensors were placed inside the chamber by using a supporting structure dedicated to this test.
	The maximum and minimum temperature during the test was 35$^{\circ}$C and $-25^{\circ}$C, respectively.
	The maximum temperature transition rate was $8^{\circ}$C per hour. 
	This slew rate is determined by the specification of the chamber, also within the requirement of the LaBr$_{3}$ producer ($<20^{\circ}$C per hour).
	A total of four cycles were performed in the test.	
	
	The test started at 09:40 UTC on 25 April 2018, and ended at 16:40 UTC on 10 May 2018, 
	with the first cycle from 11:46 on 25 April to 23:32 on 27 April, 2nd cycle from 23:22 on 1 May to 18:40 on 2 May, 
	3rd cycle from 7:15 on 3 May to 20:47 on 4 May, and 4th cycle from 22:53 on 8 May to 18:17 on 9 May.
	To prevent dew condensation and performance deterioration of the LaBr$_{3}$ crystals, the chamber was filled with 99.995\%-purity dry nitrogen gas. 
	As shown in the right panel of Figure~\ref{fig:cycle_geo}, an $^{88}$Y radiation source was used during the test and irradiated the three sensors from the side of the chamber. 
	Energy spectra of each sensor were acquired approximately every 30 min, and the gain variation was analyzed by tracing the center of the 898-keV absorption peak of $^{88}$Y in the LaBr$_{3}$ crystals.
	The temperature was measured by a platinum resistance temperature sensor PT-100 (Lake Shore Cryotronics Inc.) attached to each sensor.

	\begin{figure}
		\begin{center}
		\begin{tabular}{c}
		\includegraphics[width=\linewidth]{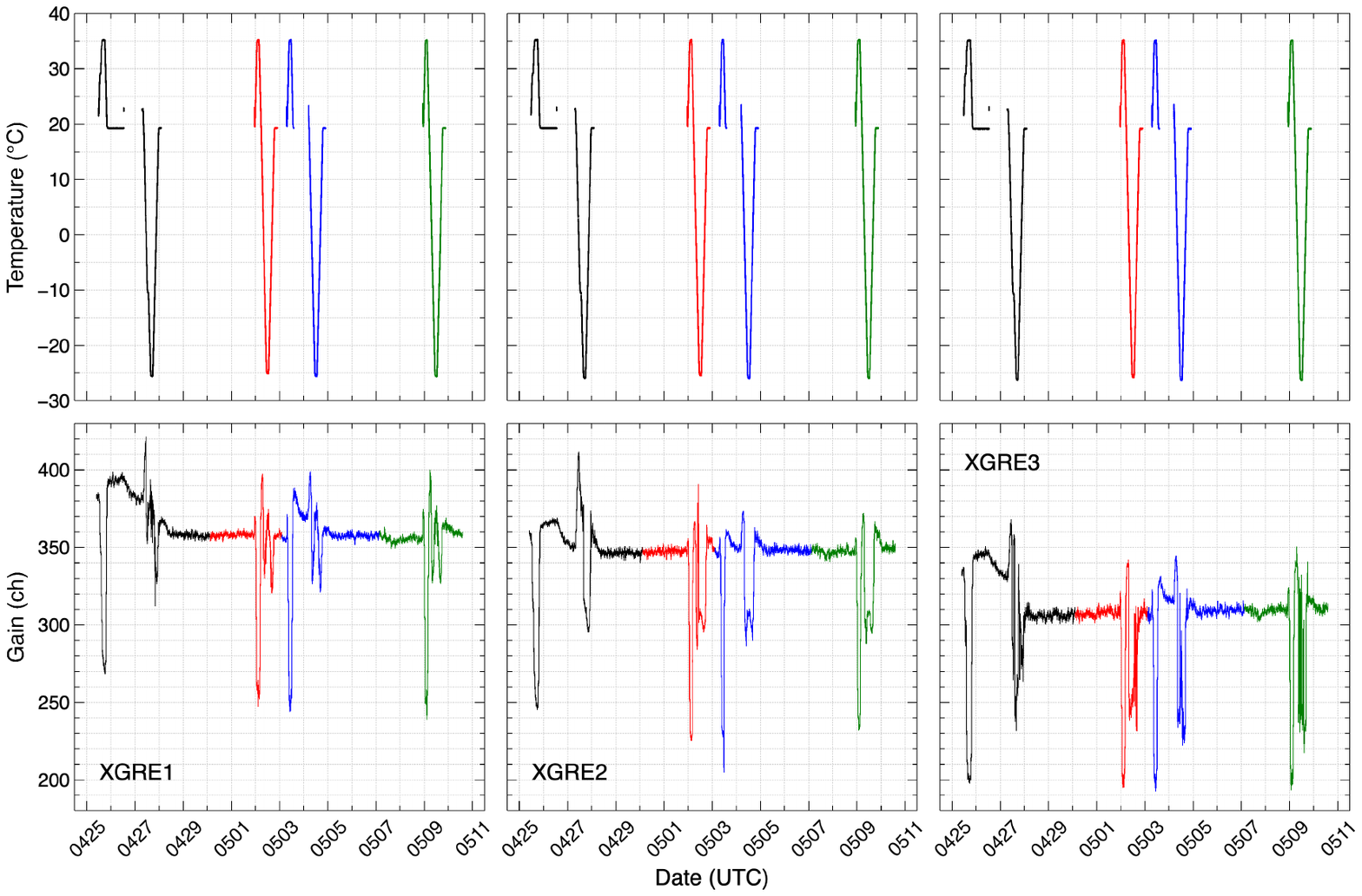}
		\end{tabular}
		\caption[Temperature and gain variation during the thermal-cycle test.]
		{Temperature (upper) and gain (lower) variation during the thermal-cycle test. Black, red, blue, and green lines indicate the first, second, third, and fourth cycles, respectively.
		The temperature profile was obtained only during the cycle operations, not during the temperature plateaux. 
		The gain is the peak channel of the 898-keV line from $^{88}$Y.\label{fig:cycle}}
		\end{center}
	\end{figure} 

	\begin{figure}
		\begin{center}
		\begin{tabular}{c}
		\includegraphics[width=\linewidth]{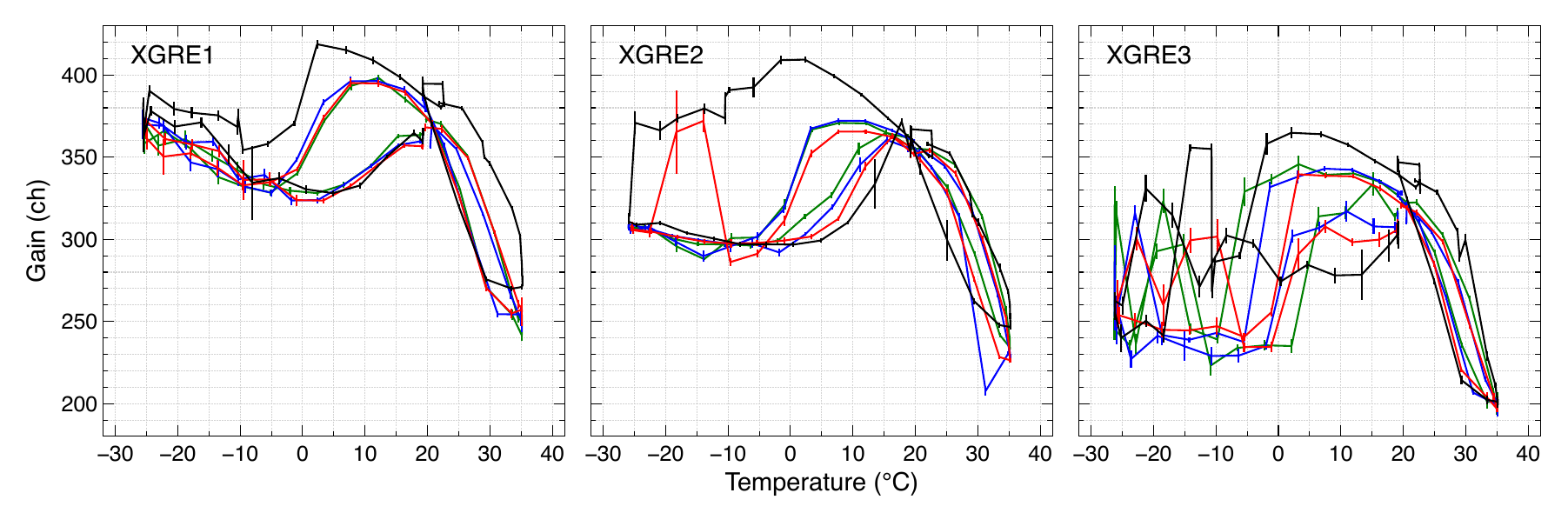}
		\end{tabular}
		\caption[The correlation between the temperature and gain profiles during the thermal-cycle test.]
		{The correlation between the temperature and gain profiles during the thermal-cycle test. Black, red, blue, and green lines indicate the first, second, third, and fourth cycles, respectively.\label{fig:cycle_cor}}
		\end{center}
	\end{figure} 

	Figure~\ref{fig:cycle} shows the temperature and gain variation of each sensor during the thermal-cycle test.
	The gain variation was tracked by fitting the total absorption peak of $^{88}$Y at 898-keV with a Gaussian function for the energy spectra acquired every 30~minutes. 
	Figure~\ref{fig:cycle_cor} shows the correlation between temperature and gain. 
	In the cycle on the high-temperature side, a negative correlation was observed, in which the gain decreased as the temperature increased. 
	In the second cycle, when the operation was stable, the gain variation was 31\%, 34\%, and 35\% for XGRE1, 2, and 3, respectively, for the temperature change from 20$^{\circ}$C to 35$^{\circ}$C.
	They correspond to the variation rate of 2.1\%~K$^{-1}$, 2.3\%~K$^{-1}$, and 2.3\%~K$^{-1}$, respectively. 
	On the other hand, in the cycle on the low-temperature side, the relationship between gain and temperature was not a monotonic function, not a one-to-one correspondence. 
	For XGRE1 and XGRE2, hysteresis was also observed in which the gain variation was different when the temperature was falling and when the temperature was rising.
	Due to the difference in plateau temperature (not shown in Figure~\ref{fig:cycle}), the gain of the first cycle during the temperature plateau is different from those of the other cycles.

	\begin{figure}
		\begin{center}
		\begin{tabular}{c}
		\includegraphics[width=0.6\linewidth]{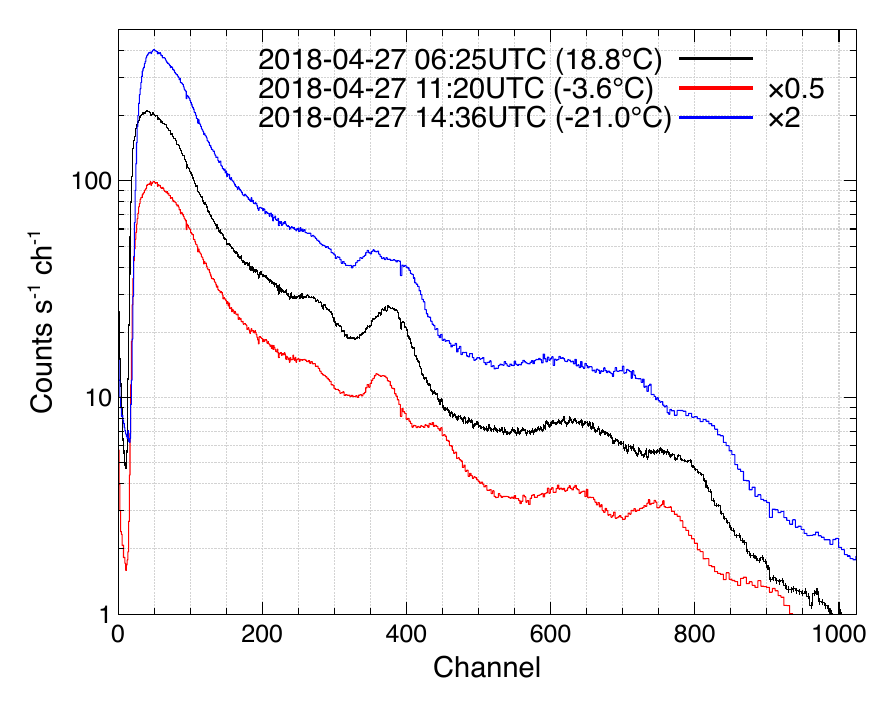}
		\end{tabular}
		\caption[Examples of energy spectra obtained by XGRE1 during the thermal-cycle test.]
		{Examples of energy spectra obtained by XGRE1 during the thermal-cycle test. The red and blue plots are shifted by a factor of 0.5 and 2, respectively, for visualization purposes.\label{fig:cycle_spec}}
		\end{center}
	\end{figure} 

	Figure~\ref{fig:cycle_spec} shows examples of the energy spectra obtained with XGRE1 on the low-temperature side of the first cycle. 
	At room temperature at 06:25UTC, we detect the 898-keV line of $^{88}$Y around 380~ch, its Compton edge around 270~ch, the 1.84-MeV line of $^{88}$Y around 770~ch, and its Compton edge around 630~ch. 
	On the other hand, at 11:20 UTC, when the temperature was falling, the 898 keV emission line was separated into two peaks. 
	At 14:36 UTC, when the temperature was falling further, the two peaks overlapped again to form a broadened emission line with a flat top.
	As described in Section~\ref{sect:det}, one sensor consisted of four detection units and two high voltage units, and one high-voltage unit served two detection units.
	The main factor of the gain variation is considered to be high-voltage variation due to temperature change. 
	The output of one sensor was the sum of the outputs of the four detection units. 
	However, if the two high voltage units have different characteristics of temperature, detection units connected to different high-voltage units could exhibit different gain variations. 	
	Therefore, the gains of the four detection units were not uniform depending on the temperature, and two peaks were generated. 
	The hysteresis of the gain-temperature relation could make hard to precisely correct the gain variation by temperature, 
	and independent control of HV units or unifying the HV units will be required for a future design.
	In the analysis, when the 898-keV emission line was separated into two, each peak was fitted with Gaussian, and the average of the two emission line centers was taken as the gain.	

\subsection{Sensor Calibration onboard Satellite}\label{sect:satcal}

\begin{figure}
		\begin{center}
		\begin{tabular}{c}
		\includegraphics[width=\linewidth]{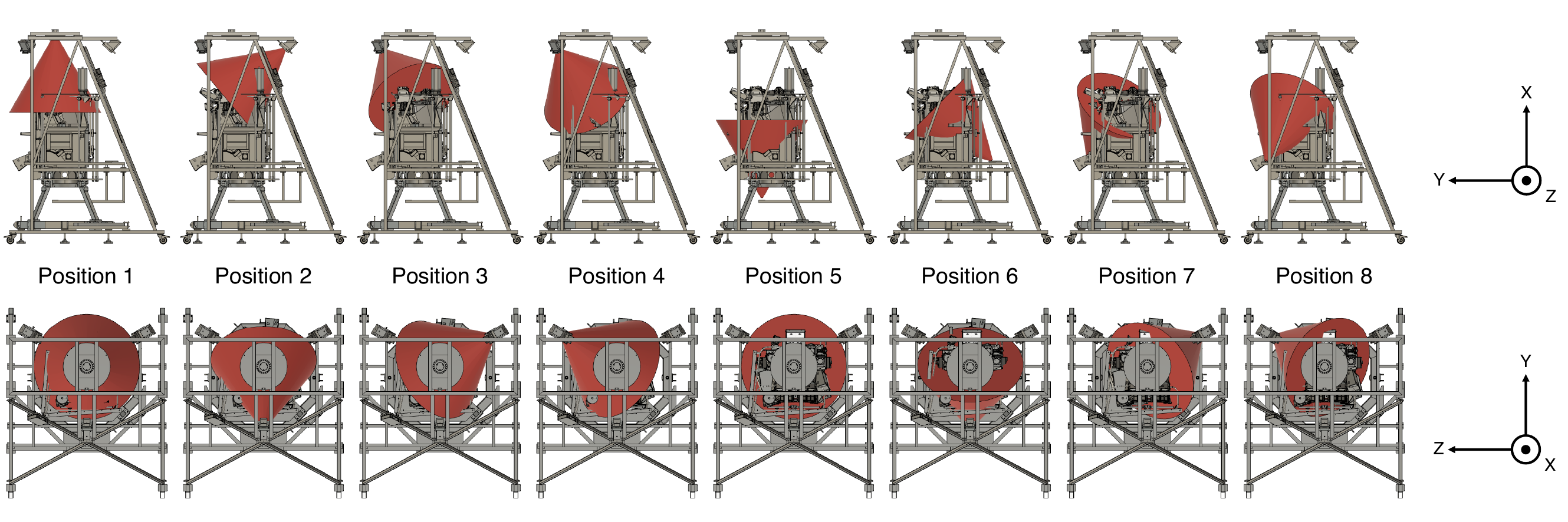}
		\end{tabular}
		\caption[Source positions for the calibration test with satellite.]
		{Source positions for the calibration test with satellite. the red cones show irradiated regions.\label{fig:cal_pos}}
		\end{center}
	\end{figure}

	After the validation and calibration tests, XGRE was delivered to CNES and installed on the TARANIS satellite. 
	The calibration test of XGRE attached to the satellite using radiation sources was performed from February to March 2020 at the CNES Toulouse Space Center. 
	As shown in Figure~\ref{fig:cal_pos}, the radiation sources irradiated the spacecraft from eight directions using a dedicated supporting structure. 
	At Position~1, gamma rays come to the detector vertically from the Nadir direction (Earthside). 
	At Positions~2--4, the azimuth angles were set to 0$^{\circ}$, 240$^{\circ}$, and 120$^{\circ}$, respectively, and the elevation angle was set to 43 $^{\circ}$. 
	At Positions~5-8, gamma rays come from the opposite side of the nadir (the space side). The elevation and azimuth angles are the same as Positions~1--4. 
	We utilized $^{137}$Cs, $^{88}$Y, $^{57}$Co, and $^{241}$Am sources.
	Their radioactivity at the time of calibration was 3.8~MBq, 3.2~MBq, 3.4~MBq, and 1.7 MBq, respectively.
	Each radiation source was enclosed in a collimator made of 95\% tungsten and 5\% iron-nickel, and attached to the supporting structure.
	The aperture angle of the collimator was 52 $^{\circ}$. We performed a total of 32 measurements with 4 sources at 8 locations and one background measurement. 
	The duration of each measurement is summarized in Table~\ref{tab:duration}. 
	Since gamma rays of $^{241}$Am were out of the energy range of XGRE2 and 3, their analysis is only included for XGRE1.

	\begin{table}
		\caption{The duration of source measurements for the calibration test with satellite.\label{tab:duration}} 
		\begin{center}       
		\begin{tabular}{| c | c c |} 
		\hline
		Source		& Positions 1--4	& Positions 5--8	\\ \hline
		$^{137}$Cs	& 0.40 hours		& 0.33 hours		\\
		$^{88}$Y		& 3.5 hours		& 0.42 hours		\\
		$^{57}$Co		& 0.26 hours		& 1.0 hour			\\
		$^{241}$Am	& 0.57 hours		& 6.0 hours		\\ \hline
		Background	& \multicolumn{2}{c|}{6.6 hours}		\\ \hline
		\end{tabular}
		\end{center}
	\end{table} 

\begin{figure}
		\begin{center}
		\begin{tabular}{c}
		\includegraphics[width=\linewidth]{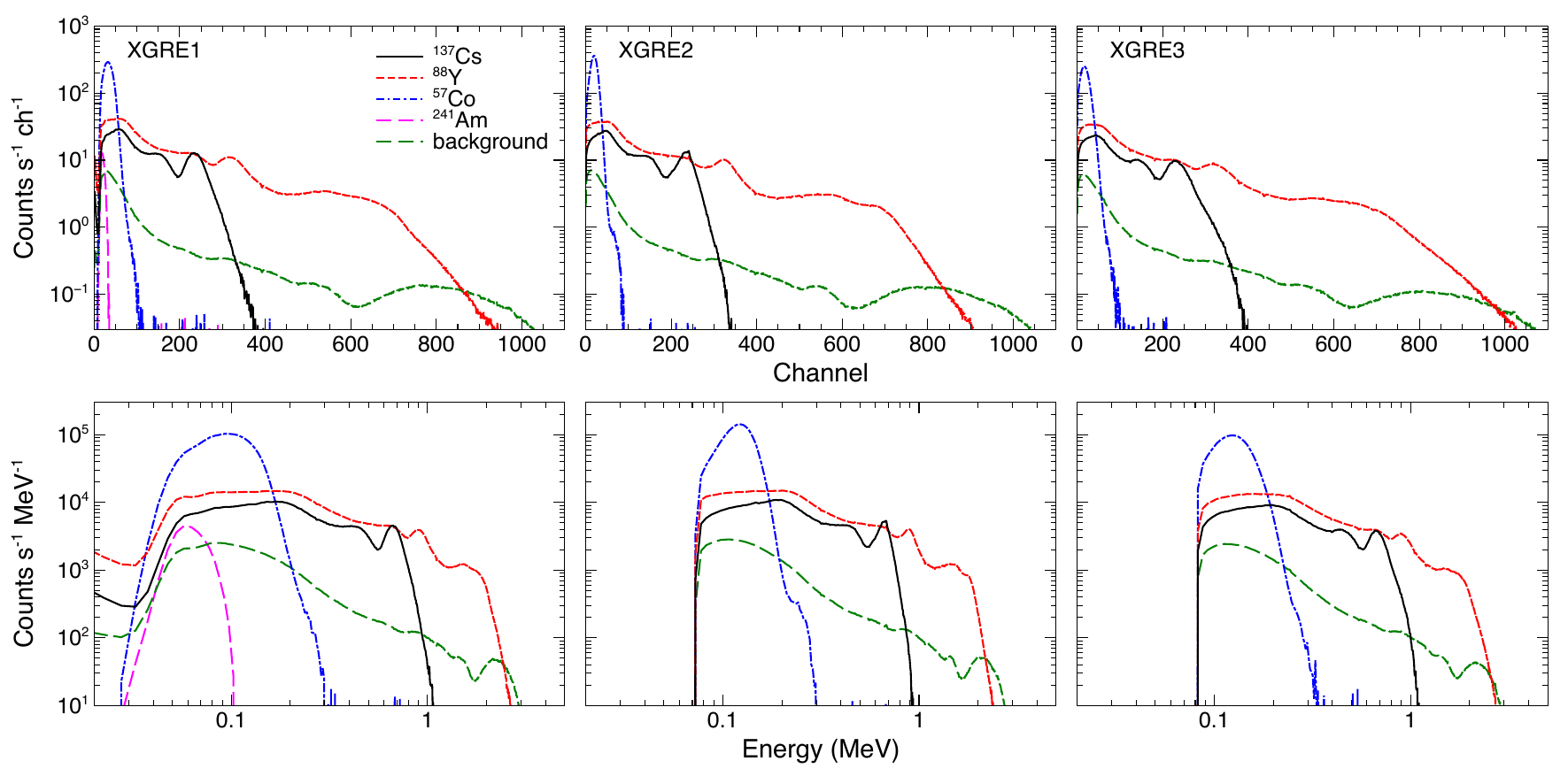}
		\end{tabular}
		\caption[Energy spectra obtained by LaBr$_{3}$ crystals during the calibration test with satellite.]
		{Energy spectra obtained by LaBr$_{3}$ crystals during the calibration test with satellite. Radiation sources were at Position~1.
		 The spectra with the radiation sources are background-subtracted. Upper: Raw spectra. Lower: Calibrated spectra.\label{fig:cal_spec}}
		\end{center}
	\end{figure}

	Figure~\ref{fig:cal_spec} shows the energy spectra measured by the LaBr$_{3}$ crystals when the radiation sources were placed at Position~1. 
	The background was subtracted for measurements with sources. 
	In the measurements with $^{241}$Am, $^{137}$Cs, and $^{88}$Y,  the 60~keV, 662~keV and 898~keV emission lines are detected, respectively. 
	With the $^{57}$Co source, a peak due to the emission lines at 122 keV and 136 keV is also found. 
	An energy calibration function was calculated by fitting these emission lines with a Gaussian function. 
	Since the two emission lines of $^{57}$Co cannot be resolved with the energy resolution of this sensor and were seen as one line, 
	the energy center was calculated as 124 keV from the branching ratio of the two lines. 
	For XGRE1, lines from $^{241}$Am, $^{137}$Cs, and $^{88}$Y are used for the calibration, 
	and $^{57}$Co is used for XGRE2 and 3 instead of $^{241}$Am.

	The calibration function was obtained as follows.
	\begin{eqnarray}
		\textrm{XGRE1:}~E = 0.0058 + 2.82\times10^{-3}\times\textrm{ch}~[\textrm{MeV}] \\
		\textrm{XGRE2:}~E = 0.0729 + 2.52\times10^{-3}\times\textrm{ch}~[\textrm{MeV}] \\
		\textrm{XGRE3:}~E = 0.0808 + 2.56\times10^{-3}\times\textrm{ch}~[\textrm{MeV}]
	\end{eqnarray}
	The energy range of each sensor was 0.04--11.6~MeV, 0.08--11.0~MeV, and 0.08--11.3~MeV for XGRE1--3, respectively. 
	XGRE1 registers energies below 0.04~MeV, but there seemed to be a cutoff below 0.04~MeV, and hence above 0.04~MeV was reliable.
	Also, the energy resolution (full width at half maximum: FWHM) at 662~keV was 20.5\%, 25.9\%, and 28.5\%, respectively.

\begin{figure}
		\begin{center}
		\begin{tabular}{c}
		\includegraphics[width=\linewidth]{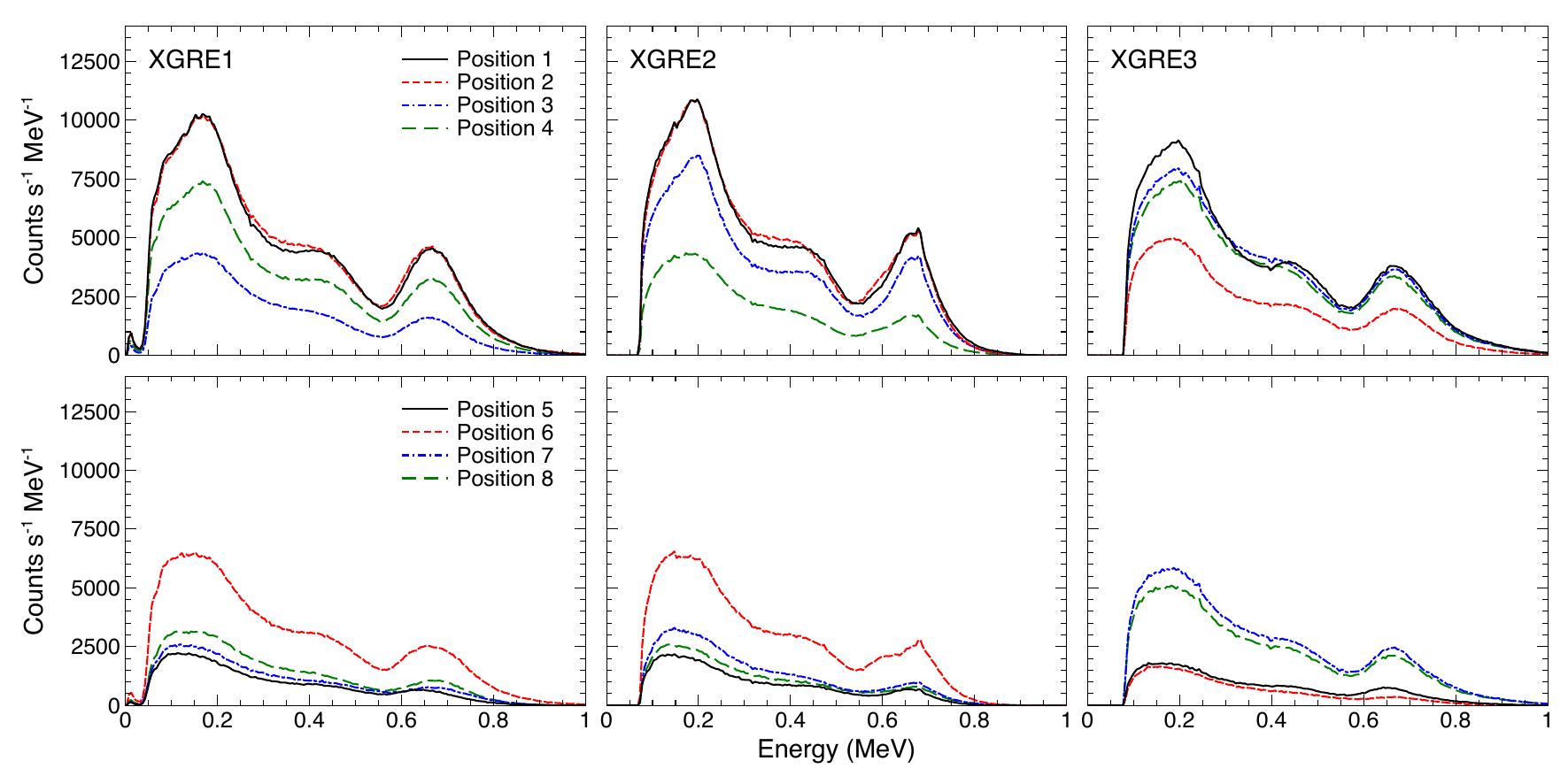}
		\end{tabular}
		\caption[Energy spectra obtained by LaBr$_{3}$ crystals with the $^{137}$Cs source at various positions.]
		{Background-subtracted energy spectra obtained by LaBr$_{3}$ crystals during the calibration test with satellite, with the $^{137}$Cs source at various positions.\label{fig:cal_spec_pos}}
		\end{center}
	\end{figure}

	Figure~\ref{fig:cal_spec_pos} shows the energy spectra of the LaBr$_{3}$ crystals with the $^{137}$Cs source at each position. 
	The intensity varies depending on the position of the radiation source. For example, in the measurement at Positions~2 and 6,
	the intensities of XGRE1 and 2 were higher than that of XGRE3. 
	These measurements are checked in Section~\ref{sect:sim} using Monte Carlo simulations.

\begin{figure}
		\begin{center}
		\begin{tabular}{c}
		\includegraphics[width=\linewidth]{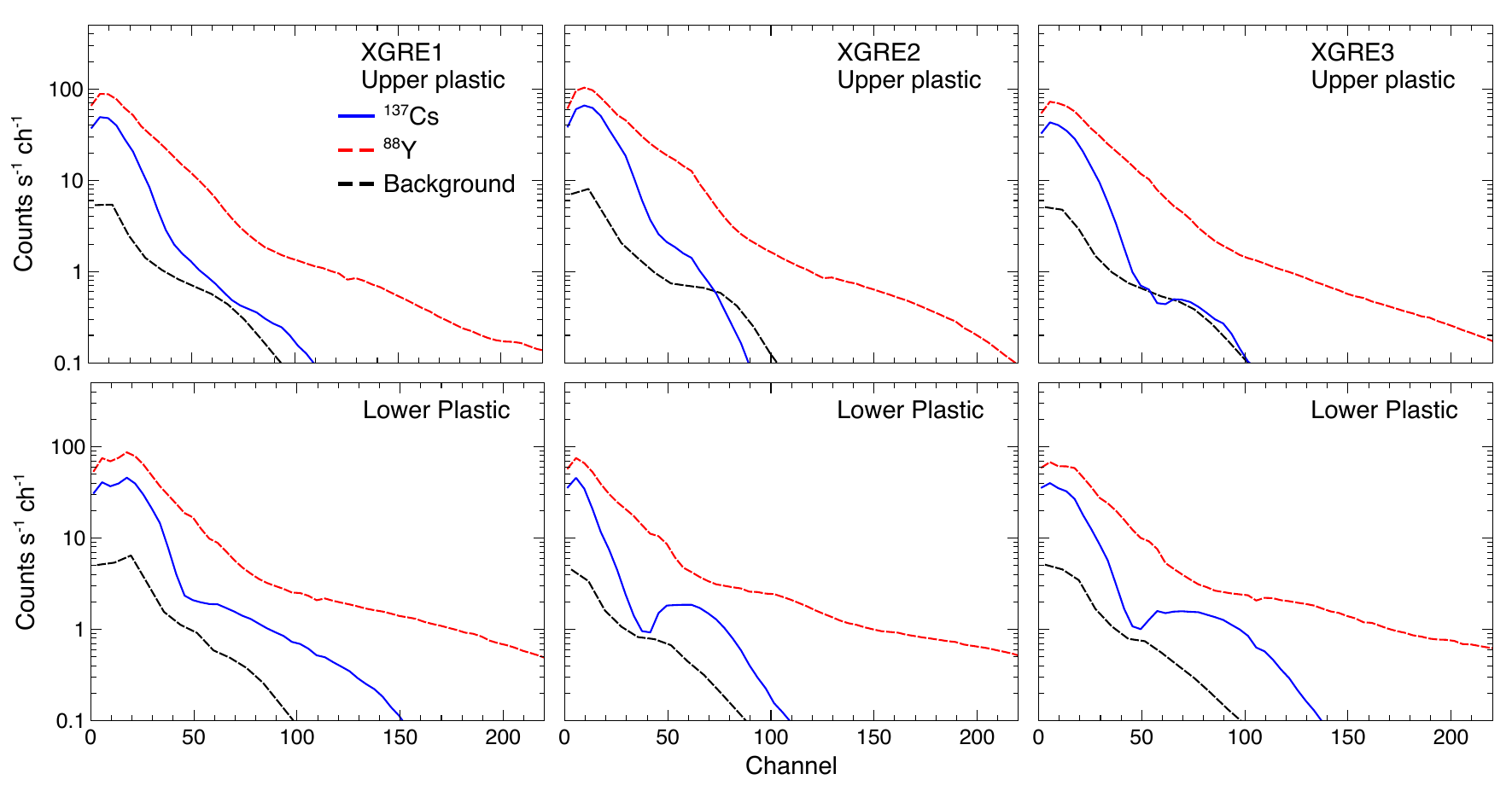}
		\end{tabular}
		\caption[Energy spectra obtained by plastic scintillators during the calibration test with satellite.]
		{Background-subtracted energy spectra obtained by plastic scintillators during the calibration test with satellite.
		The radiation sources were at Position~1.\label{fig:cal_spec_pla}}
		\end{center}
	\end{figure}

	Figure \ref{fig:cal_spec_pla} shows the energy spectra obtained with plastic scintillators when the $^{137}$Cs or $^{88}$Y source was placed at Position~1. 
	Since plastic scintillators have small cross sections for photoabsorption, energy calibration is usually performed using a Compton edge. 
	However, due to the moderate energy resolution, the plastic scintillators of XGRE barely detect the Compton edge due to the 662-keV line of $^{137}$Cs and the 898-keV line of $^{88}$Y. 
	Therefore, accurate energy calibration of the plastic scintillator was difficult with the gamma-ray sources and a specific method had to be developed to calibrate the upper plastic scintillator (see Laurent et al.\cite{Laurent_2023}).
	Note that a valley-like structure in the energy spectra of lower plastic scintillators with $^{137}$Cs is confirmed around channel~40, which seems not to be explained by general Compton-scattering features.

\section{Performance Verification with Monte Carlo Simulations}\label{sect:sim}

	In this section, we perform Monte Carlo simulations with a satellite model and validate the simulation results by comparing them with the measurement results obtained in the satellite calibration test. 
	We also evaluate the characteristics of the XGRE, such as effective area, angular resolution, and electron detection, using the verified simulation model. 
	Monte Carlo simulations were performed with Geant4 version~4.10.3\cite{Agostinelli_2003,Allison_2006,Allison_2016} and CADMesh version~1.1\cite{Poole_2012}. 
	The satellite/detector models implemented in Geant4 are based on those used in Sarria et al.\cite{Sarria_2017}

\subsection{Verification of Geant4 model}\label{sect:verification}

\begin{figure}
		\begin{center}
		\begin{tabular}{c}
		\includegraphics[width=0.8\linewidth]{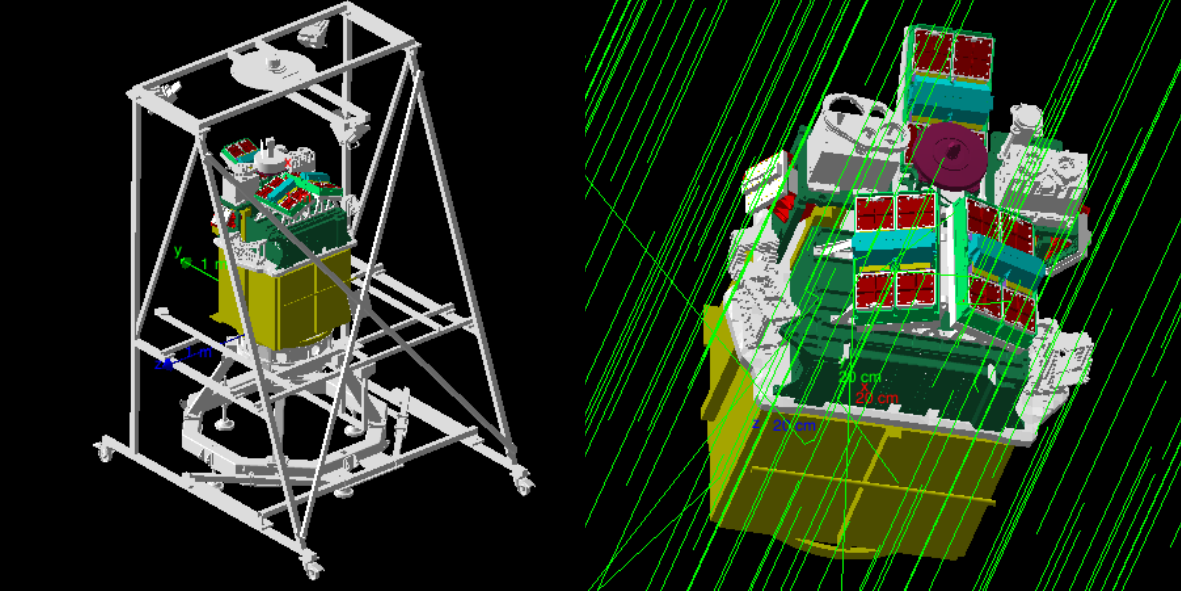}
		\end{tabular}
		\caption[Schematics of Geant4 simulations.]
		{Schematics of Geant4 simulations. Left: Simulation for the on-ground calibration test. Right: Simulation in the space environment.\label{fig:sim_geo}}
		\end{center}
	\end{figure}

	We constructed a model simulating the calibration test with the satellite conducted at CNES, then performed Monte Carlo simulations.
	The simulation model is shown on the left of Figure~\ref{fig:sim_geo}.
	The satellite/detector, a supporting structure for the satellite, a support for attaching the radiation sources, 
	and a collimator for the radiation sources were implemented in the simulation space where the atmosphere was also implemented.
	Here, we generated gamma rays from the $^{137}$Cs and $^{88}$Y sources at Positions ~1--4, calculated the energy deposits in the LaBr$_{3}$ crystal of each sensor, and extracted the energy spectra.
	In the simulated spectrum, the energy resolution at FWHM on the calibration test was convoluted as follows.
	\begin{eqnarray}
		\Delta E = \Delta E_{662} \times \left( \frac{E}{\rm 0.662~MeV} \right)^{0.5}
	\end{eqnarray}
	$\Delta E_{662}$ is the energy resolution at 0.662~MeV, which is 20.5\%, 25.9\%, and 28.5\% for XGRE1--3, respectively. 

	Here we assume that these energy resolutions are due to a reduced LaBr3 optical efficiency, consequence of the limited surface area of the optical contact
	between LaBr$^{3}$ crystals and PMTs (scintillation light losses), rather than electronic noises.
	The normalization of the simulation results was calculated from the number of gamma rays generated in the simulation, 
	the radioactivity of the radiation sources during the calibration test, and the measurement time. 
	In each simulation, $5\times10^{8}$ gamma rays were generated.

\begin{figure}
		\begin{center}
		\begin{tabular}{c}
		\includegraphics[width=\linewidth]{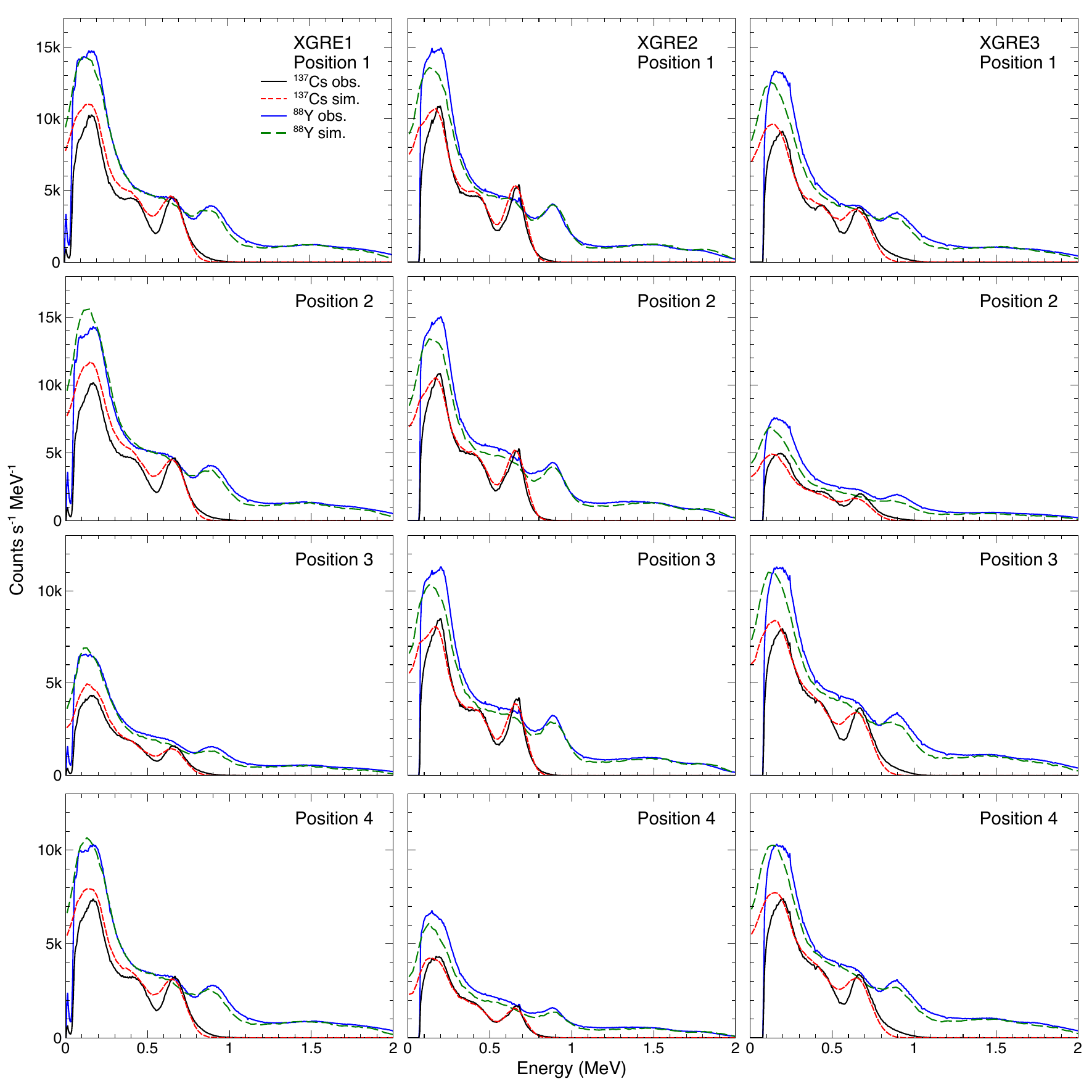}
		\end{tabular}
		\caption[Comparison of energy spectra obtained by calibration and simulation.]
		{Comparison of energy spectra obtained by the calibration test with satellite and by Monte Carlo simulations.
		Solid and dashed lines indicate measurement and simulation spectra, respectively.\label{fig:sim_spec}}
		\end{center}
	\end{figure}

	Figure~\ref{fig:sim_spec} compares the simulated spectra with the spectra obtained in the calibration test. 
	At any source type and source position, the emission line from the sources and their Compton scattering component is generally reproduced above 0.2~MeV by the simulation spectra. 
	In the energy band below 0.2~MeV, the measurement results tend to exceed the simulation results, especially those with the $^{88}$Y source. 
	The low-energy component is highly susceptible to scattering by materials around the satellite and detector, and hence it may cause the excess below 0.2~MeV
	(note that no floor nor walls was implemented in the simulation).
	Therefore, the simulation model is validated above 0.2~MeV, and has an uncertainty of $\sim$15\% at a maximum below 0.2~MeV.

\subsection{Effective areas for X-rays and gamma rays}\label{sect:effective}

\begin{figure}
		\begin{center}
		\begin{tabular}{c}
		\includegraphics[width=0.5\linewidth]{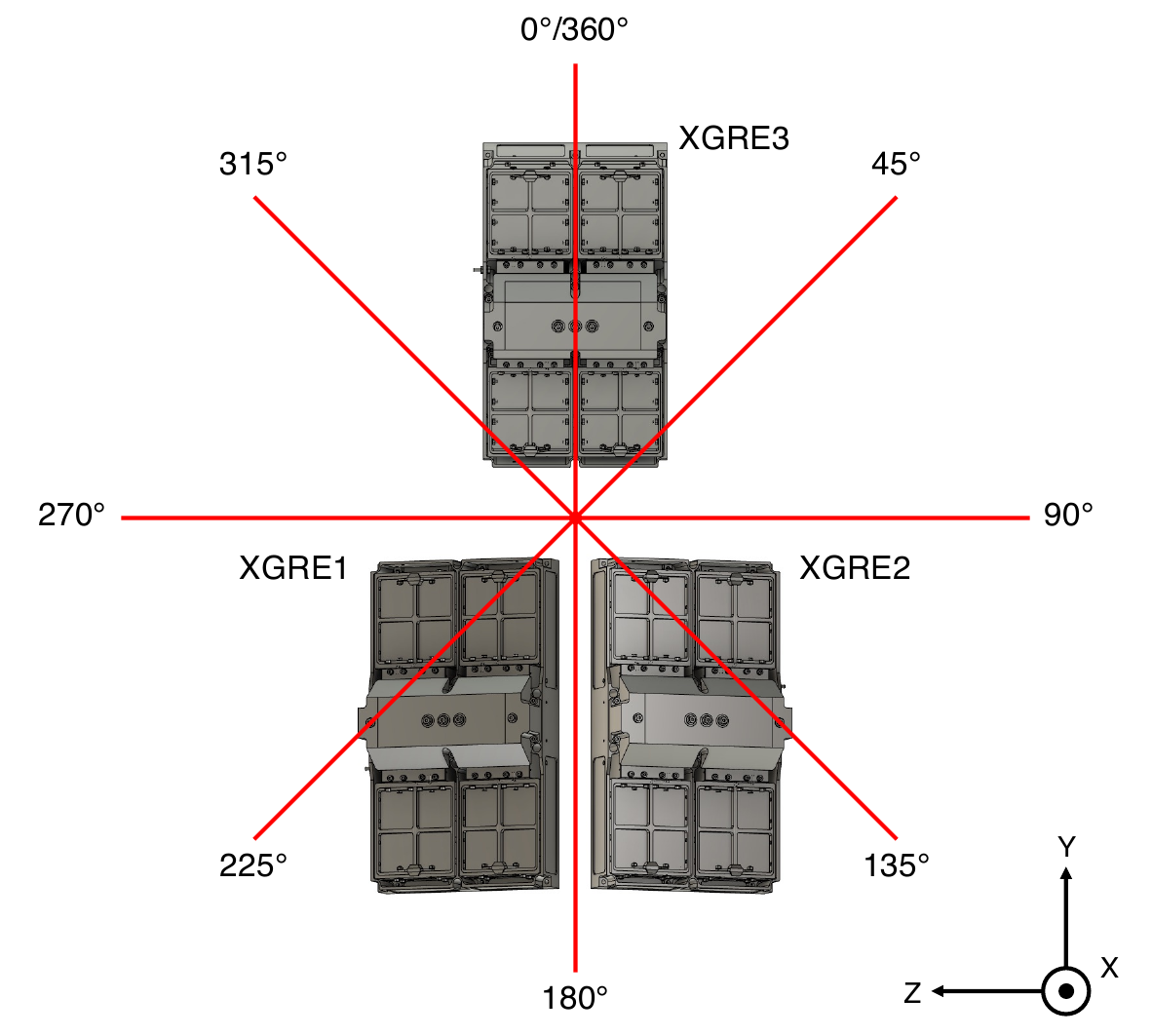}
		\end{tabular}
		\caption[Definition of azimuth angle.]
		{Definition of azimuth angle.\label{fig:azimuth}}
		\end{center}
	\end{figure}

	Using the simulation model verified in Subsection~\ref{sect:verification}, we calculated the effective area for X-rays and gamma-rays of XGRE. 
	Here the effective area is defined as an area for depositing at least the threshold energy in the detector. The simulation model is shown on the right panel of Figure~\ref{fig:sim_geo}. 
	We prepared the satellite model in space orbit and generated photons in a vacuum. 
	In the simulation, 10$^{9}$ photons from 0.02~MeV to 10~MeV were emitted from a disk-shaped region with a radius of 1~m (a fluence of $3.18\times10^{4}$~photons~cm$^{-2}$), 
	and the effective area was calculated from the energy deposits in the LaBr$_{3}$ crystal. 
	In addition, two elevation angles of 90$^{\circ}$ (the nadir direction) and 60$^{\circ}$ were set for the particle generation direction, 
	and simulations were performed for eight azimuth angles when the elevation angle was 60$^{\circ}$. 
	The definition of azimuth angles is shown in Figure~\ref{fig:azimuth}.

\begin{figure}
		\begin{center}
		\begin{tabular}{c}
		\includegraphics[width=\linewidth]{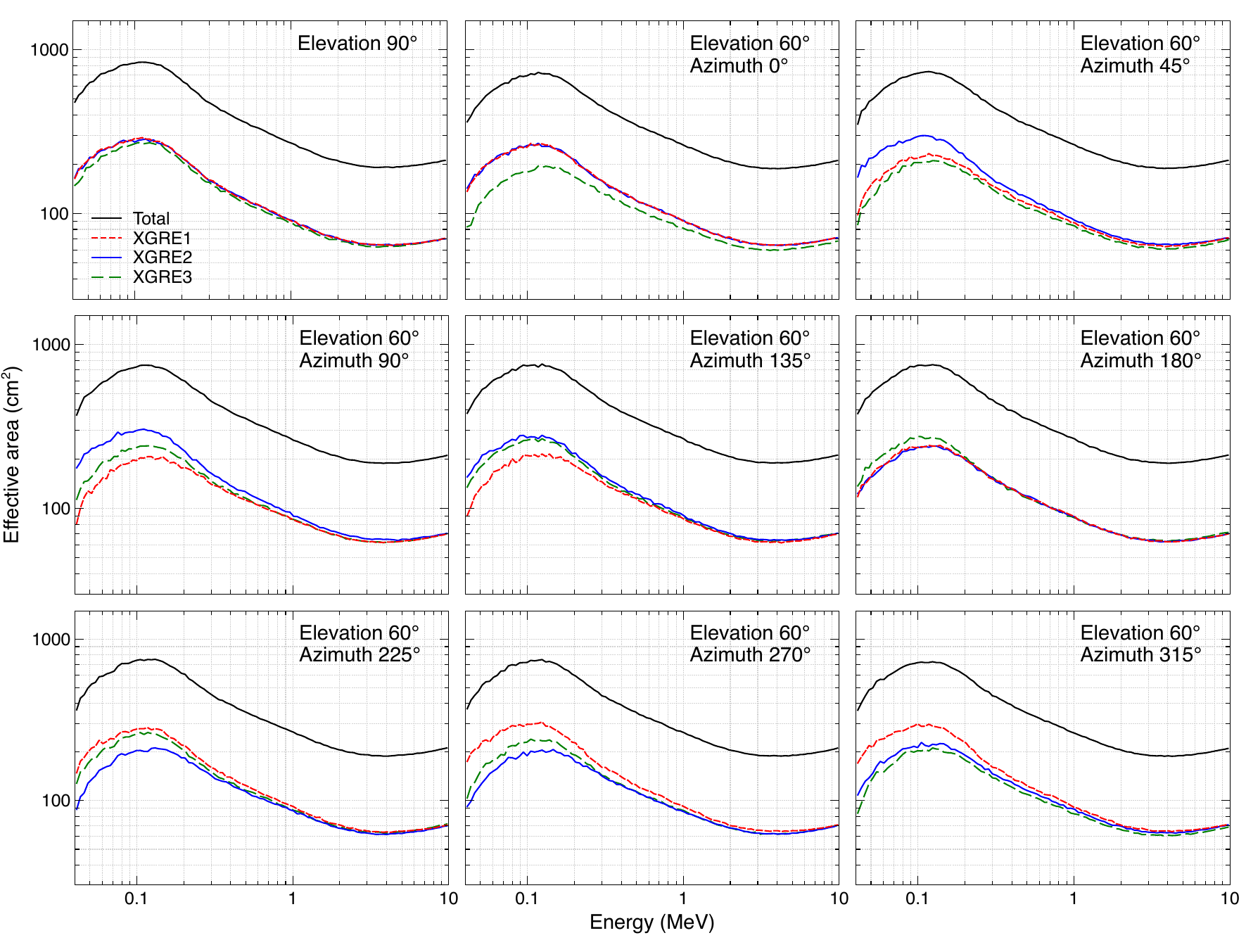}
		\end{tabular}
		\caption[Effective areas of X-rays and gamma rays for various elevation and azimuth angles.]
		{Effective areas of X-rays and gamma rays for various elevation and azimuth angles.
		The 90$^{\circ}$ elevation angle means that photons come from the nadir direction.
		The definition of azimuth angle is in Fig~\ref{fig:azimuth}.\label{fig:area}}
		\end{center}
	\end{figure}

	Figure~\ref{fig:area} shows the calculated effective area. At an elevation angle of 90$^{\circ}$, the effective area reaches the maximum at 0.1 MeV, 838~cm$^{2}$ in total for the three sensors. 
	Since the geometric area of the LaBr$_{3}$ scintillators at the elevation angle of 90$^{\circ}$ is 850~cm$^{2}$, the detection efficiency is 98.6\% at 0.1~MeV. 
	The effective area reaches the minimum at 3.5~MeV, 191~cm$^{2}$ in total for the three sensors. At this time, the detection efficiency is 22.8\%.
	The result is consistent with the previous work.\cite{Sarria_2017}
	In the band below 0.3~MeV at an elevation angle of 60$^{\circ}$, the effective area of the sensors varies depending on the azimuth angle.

	Then we simulated an energy spectrum of a TGF detected by XGRE.
	Briggs et al.\cite{Briggs_2010} reported that a typical fluence of TGFs in an energy range of 0.2--40~MeV $F_{0.2-40~{\rm MeV}}$, detected by Fermi, was 0.7~photons~cm$^{-2}$.
	Here we assume a typical energy spectrum of TGFs\cite{Dwyer_2012b} $dN/dE$
	\begin{equation}\label{eq:tgf}
		\frac{dN}{dE} \propto E^{-1} \exp\left(-\frac{E}{\rm 7.3~MeV}\right),
	\end{equation}
	where $E$ is the energy of gamma rays. A fluence $F_{0.04-10~{\rm MeV}}$ in the observation range of XGRE1 0.04--10~MeV is then
	\begin{equation}
		F_{0.04-10~{\rm MeV}} = F_{0.2-40~{\rm MeV}}  \times \frac{\int^{\rm 10~MeV}_{\rm 0.04~MeV} E^{-1}
		\exp\left(-\frac{E}{\rm 7.3~MeV}\right) dE}{\int^{\rm 40~MeV}_{\rm 0.2~MeV} E^{-1} \exp \left(-\frac{E}{\rm 7.3~MeV}\right) dE}
		= 1.06~\textrm{photons~cm}^{-2}.
	\end{equation}
	This is a typical fluence at Fermi's orbital altitude of 550~km.
	Assuming that TGFs are typically produced at an altitude of 15~km, the fluence at the orbital altitude of TARANIS, 700~km, is
	\begin{equation}
		1.06 \times \left( \frac{550-15~\textrm{km}}{700-15~\textrm{km}} \right)^{2} = 0.65~\textrm{photons~cm}^{- 2}.
	\end{equation}

\begin{figure}
		\begin{center}
		\begin{tabular}{c}
		\includegraphics[width=0.5\linewidth]{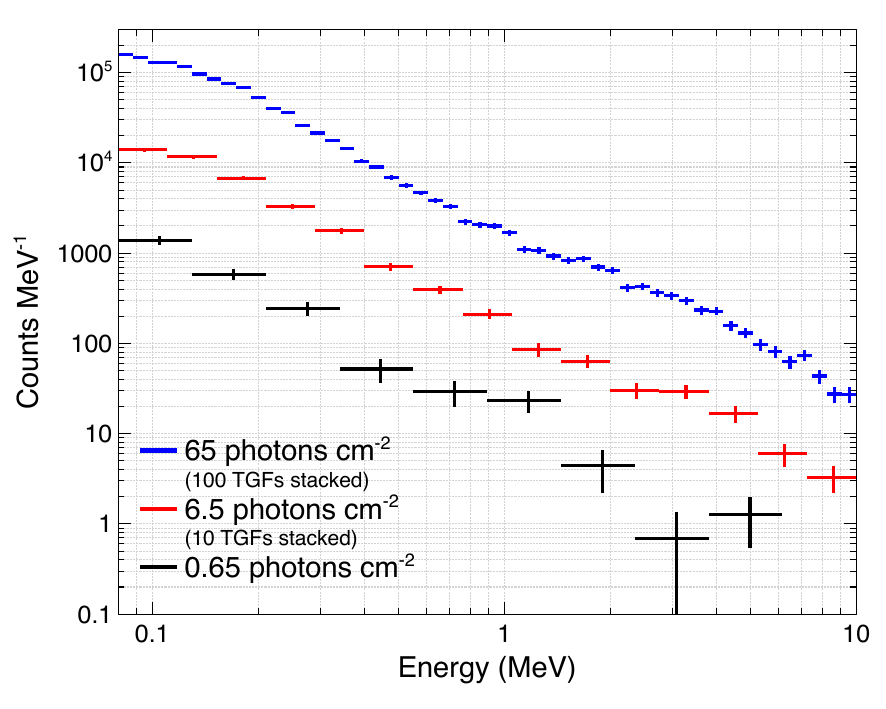}
		\end{tabular}
		\caption[Simulation spectra of TGFs by XGRE with various photon fluences.]
		{Simulation spectra of TGFs by XGRE with various photon fluences. 
		The fluence of 0.65~photons~cm$^{-2}$ is an expected one of a typical TGF (see the main text).\label{fig:tgf_spec}}
		\end{center}
	\end{figure}

	Figure~\ref{fig:tgf_spec} shows the simulated TGF spectrum assuming detection by XGRE. 
	We performed simulations with the the energy spectrum following Equation~\ref{eq:tgf}, an elevation angle of 90$^{\circ}$,
	and with three fluences of 0.65, 6.5 and 65~photons~cm$^{-2}$ (1~TGF, 10~TGFs stacked, 100~TGFs stacked, respectively). At the typical TGF fluence, 0.65~photons~cm$^{-2}$, 
	both the number of photons and the effective area of XGRE were small in the band above 5-MeV, and few photons were detected. 
	Spectral analysis of a single TGF would be available only on the low-energy side.
	On the other hand, by performing a stacking analysis that adds spectra of multiple TGFs, it would be possible to discuss the power-law indexes and cut-off energies.

\subsection{Effective areas for electrons}\label{sect:ele_effective}

	XGRE identified electrons coming from the nadir direction when the upper plastic and the LaBr$_{3}$ scintillators among the three ones in a detection unit detected energy deposits simultaneously;
	non-charged particles (mainly X-rays and gamma rays) when only one of the three scintillators was hit; 
	electrons from space when the LaBr$_{3}$ and the bottom plastic scintillators were hit simultaneously;
	high-energy charged particles such as protons and electrons with energies beyond the detection range when the three layers were hit simultaneously.
	When electrons are detected, their kinetic energy is deposited in the upper plastic and LaBr$_{3}$ scintillators, and the amounts of deposits depend on the energy of the incident electrons.
	Therefore, by generating electrons in a simulation and calculating the energy deposits in each scintillator, 
	the incident energy of the electrons can be estimated from the energy deposits in the scintillators.

	In the simulation, we used the same setup as in Subsection~\ref{sect:effective} (the right panel of Figure~\ref{fig:sim_geo}),
	and generated 1--10-MeV electrons with a fluence of $3.18\times10^{4}$~electrons~cm$^{-2}$ (10$^{9}$ electrons in a 1~m radius disk) and an elevation angle of 90$^{\circ}$.	
	In a single particle run, electron detection was identified when the upper plastic and LaBr$_{3}$ scintillators detected energy deposits exceeding their energy threshold.	
	For the LaBr$_{3}$ crystals, the energy threshold was 0.04~MeV for XGRE1 and 0.08~MeV for XGRE2 and 3 as determined by the satellite calibration.
	While the exact threshold for plastic scintillators has not been determined due to the difficulty of energy calibration, it was assumed to be 0.2~MeV.
	In real TEB cases, electrons are bound by Earth's magnetic field, and have various pitch angles.
	Therefore, the represent simulation with parallel electron beams is limited, and further investigation is needed to apply this to realistic TEBs (beyond the scope of the present paper).

	\begin{figure}
		\begin{center}
		\begin{tabular}{c}
		\includegraphics[width=0.9\linewidth]{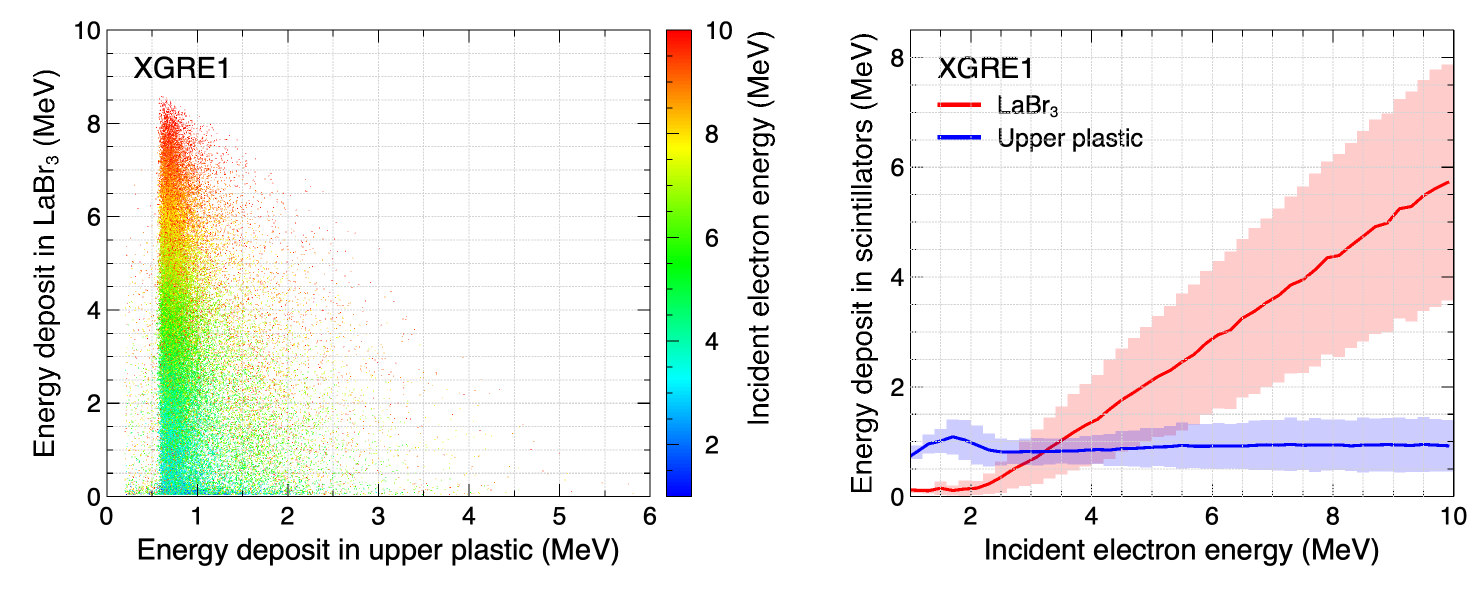}
		\end{tabular}
		\caption[Simulated behaviors of electrons in XGRE.]
		{Simulated behaviors of electrons in XGRE. Left: A scatter plot for the relation between energy deposits in LaBr$_{3}$ and upper plastic scintillators. The plot colors indicate initial electron energies.
		Right: Relations between incident electron energy and energy deposits in scintillators. The translucent areas show statistical error at one standard deviation.\label{fig:ele_scat}}
		\end{center}
	\end{figure} 

	The left panel of Figure~\ref{fig:ele_scat} shows a scatter plot of the relationship between the energy deposit in the upper plastic and the LaBr$_{3}$ scintillator for electron detection by XGRE1. 
	The color of the plot corresponds to the incident electron energy. The distribution of energy deposit in the plastic scintillators is concentrated around 0.6--1.2~MeV. 
	On the other hand, energy deposits in LaBr$_{3}$ increase when incident electron energy increases. 
	It is noted that the energy deposits in both the LaBr$_{3}$ and upper plastic scintillators show variations.
	Electrons continuously react with matter and deposit their kinetic energy through numerous scatterings called straggling. 
	Since straggling is a stochastic process, the paths in the scintillator are slightly different even for electrons with the same incident energy, resulting in a difference in the energy deposit. 
	To determine that the XGRE detected an electron, the electrons must penetrate the upper plastic scintillator, and deposit all the rest kinetic energy in LaBr$_{3}$. 
	Therefore, the energy deposit in the plastic scintillator is almost constant (determined by the thickness of the scintillator), and the remaining energy is deposited in LaBr$_{3}$.

	The right panel of Figure~\ref{fig:ele_scat} shows the relation between incident electron energy and average energy deposits in each scintillator. 
	In addition, the semi-transparent regions indicate the variation of energy deposits at one standard deviation.
	The average energy deposit in the plastic scintillators is around 1~MeV for the entire range of incident electron energy. 
	Energy deposits in the upper plastic are at maximum with incident electron energies between 1.5~MeV and 2~MeV. 
	This is due to the large energy deposit per unit length in lower kinetic energies, according to the Bethe formula. 
	For the LaBr$_{3}$ scintillators, the energy deposit is a linear function of the incident energy above 3 MeV. 
	Therefore, for electrons above 3 MeV, the incident energy of electrons can be estimated from the energy deposit at LaBr$_{3}$. 
	However, we must consider statistical variations in practice. For example, when the energy deposit in LaBr$_{3}$ is 2~MeV, 
	the estimated incident electron energy is 3.8--6.7 MeV with one standard deviation, an uncertainty of $\sim$30\%. 
	In addition, the energy resolution also affects the measurement accuracy of the energy deposits in LaBr$_{3}$, which is considered to cause a larger uncertainty. 
	Figure~\ref{fig:ele_scat} shows only XGRE1, but we obtained the same results for XGRE2 and 3.

	\begin{figure}
		\begin{center}
		\begin{tabular}{c}
		\includegraphics[width=\linewidth]{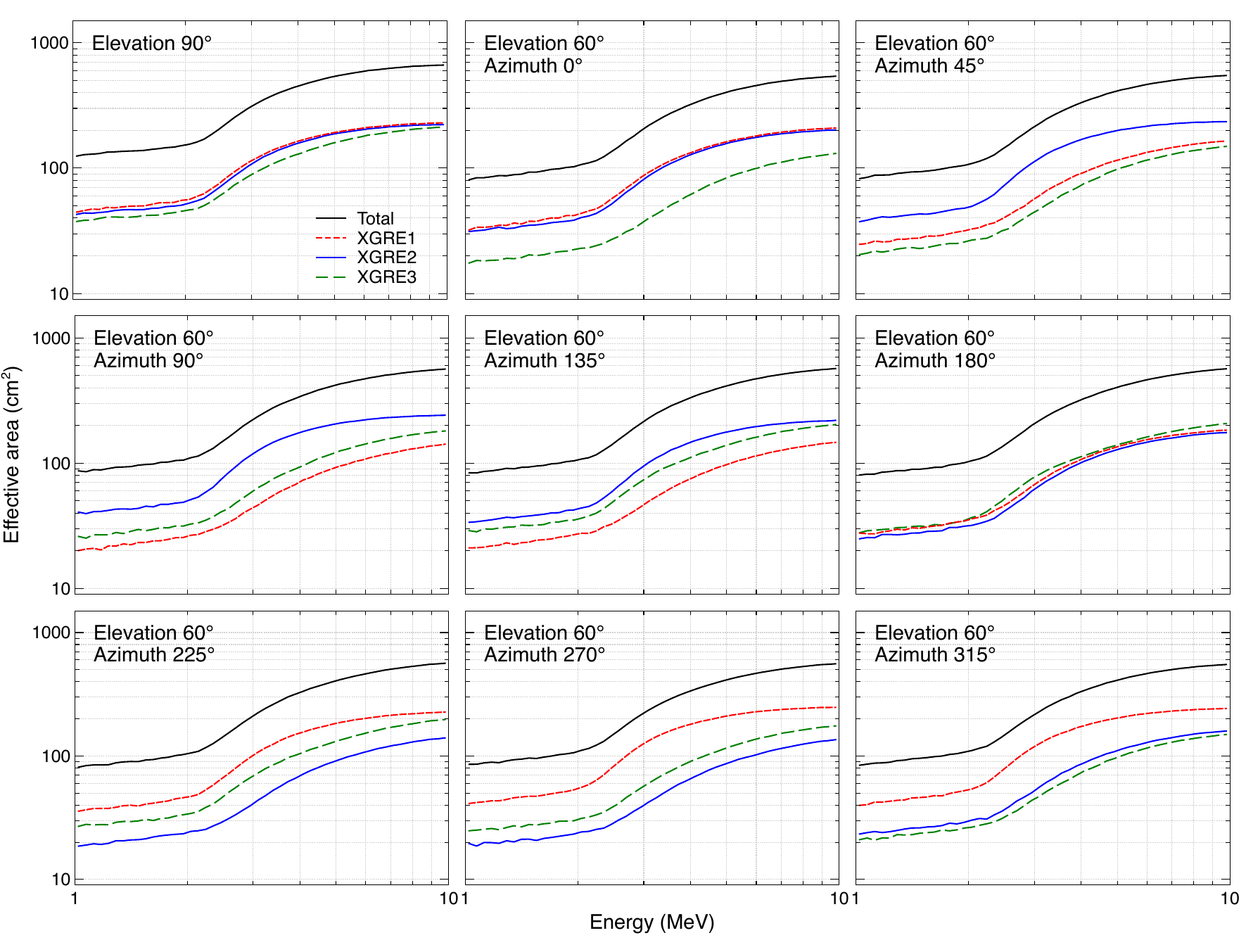}
		\end{tabular}
		\caption[Effective areas of electrons for various elevation and azimuth angles.]
		{Effective areas of electrons for various elevation and azimuth angles, as the same condition as Fig \ref{fig:area}.\label{fig:ele_area}}
		\end{center}
	\end{figure} 

	Figure~\ref{fig:ele_area} shows the effective area of XGRE for electrons. 
	The criteria for electron detection are the same as the analysis above. 
	At an elevation angle of 90$^{\circ}$, the total effective area of the three sensors is 155 cm$^{2}$ for electrons with an incident energy of 2~MeV, 
	but it monotonically increases with the incident electron energy, and It is 664~cm$^{2}$ for electrons with an incident energy of 10~MeV. 
	The effective areas at an elevation angle of 60$^{\circ}$ were also calculated in the same way as Figure~\ref{fig:area}. 	
	In the case of electrons, the change in effective area with azimuth angle is observed in the entire energy range of 1--10~MeV, 
	and the amount of changes by an incident angle is higher than that of photons.

\subsection{Angular Resolution}\label{sect:res_angle}

	As shown in Figure~\ref{fig:dimension}, the three sensors of XGRE were mounted at angles, 
	and the effective area changes with elevation and azimuth angles of incident particles on the low-energy side, as shown in Figure~\ref{fig:area}.	
	By utilizing this characteristic, the direction of arrival of gamma rays can be measured from the ratio of photon counts of the three sensors.
	We verified this directional resolution by Monte Carlo simulations. The simulations had the same setup as in Subsection~\ref{sect:effective} (the right of Figure \ref{fig:sim_geo} right). 
	Here, the incident particles were gamma rays following the TGF spectrum (Equation~\ref{eq:tgf}), with a fluence of $3.18\times10^{3}$~photons~cm$^{-2}$ in 0.04--10~MeV. 
	A total of 109 simulations were performed for four elevation angles of 90$^{\circ}$, 75$^{\circ}$, 60$^{\circ}$, 45$^{\circ}$, 
	and 36 azimuth angles (10$^{\circ}$ interval) for the three elevation angles excluding 90$^{\circ}$. 
	The thresholds of the LaBr$_{3}$ crystals were set to 0.08~MeV for all the three sensors.

	\begin{figure}
		\begin{center}
		\begin{tabular}{c}
		\includegraphics[width=\linewidth]{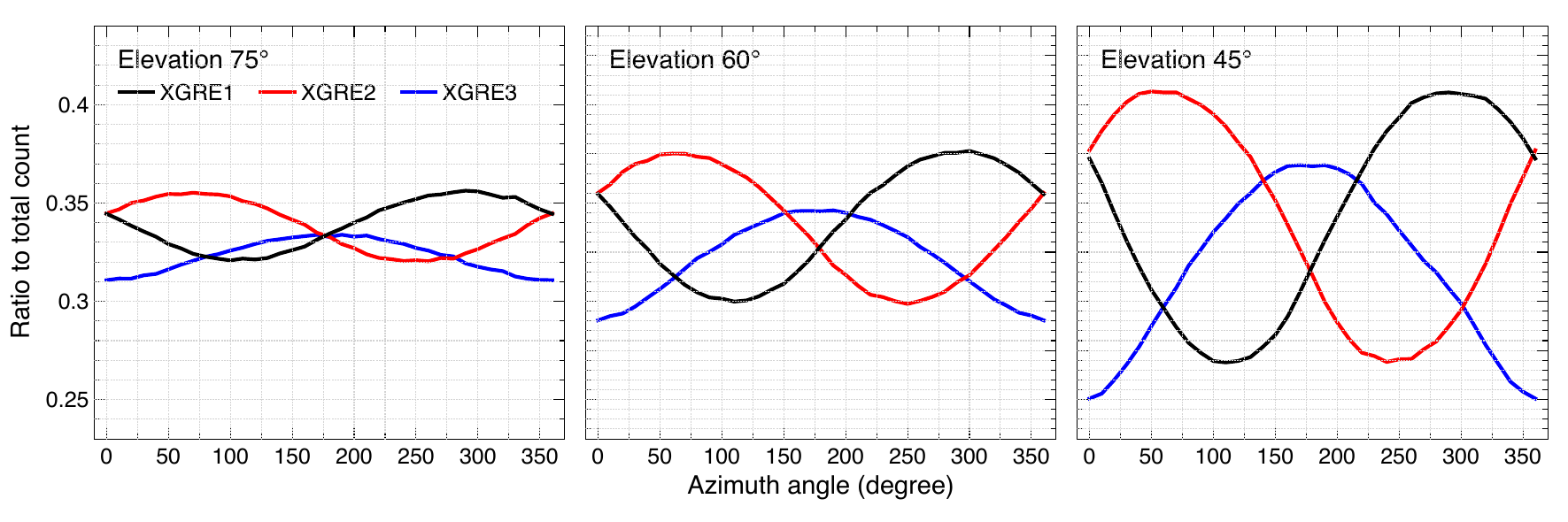}
		\end{tabular}
		\caption[The relation between azimuth angles of gamma rays and photon counts by XGRE.]
		{The relation between azimuth angles of gamma rays and photon counts by XGRE. The photon counts are normalized by the total count of XGRE1--3 at each azimuth angle.\label{fig:ang_var}}
		\end{center}
	\end{figure} 

	\begin{figure}
		\begin{center}
		\begin{tabular}{c}
		\includegraphics[width=0.6\linewidth]{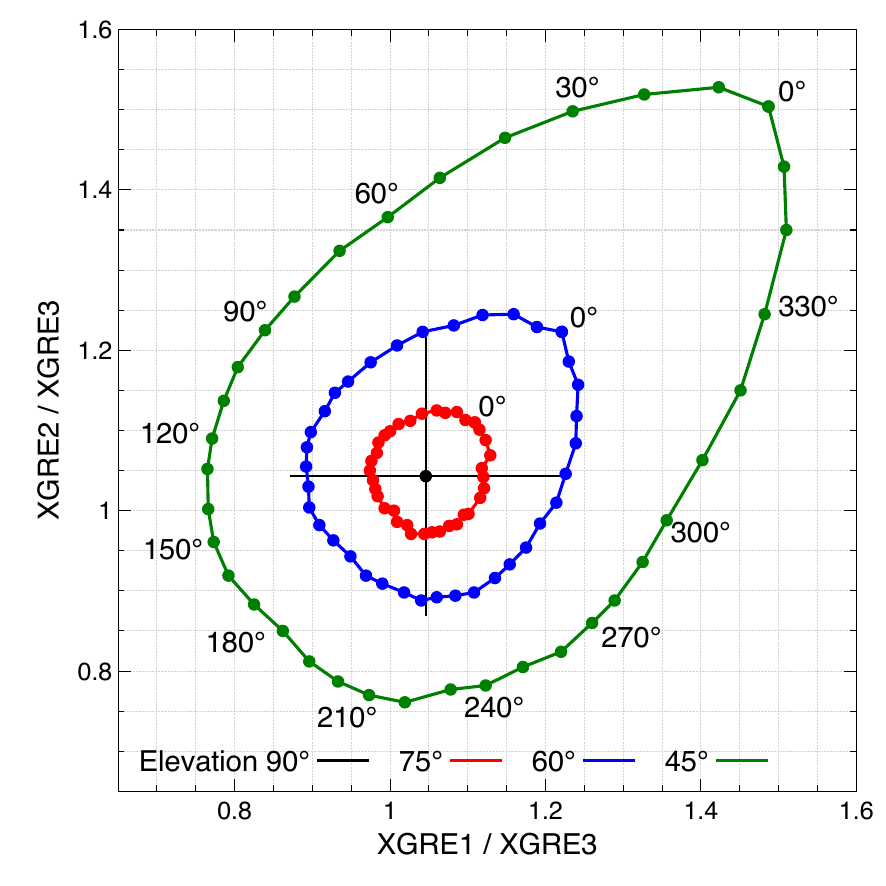}
		\end{tabular}
		\caption[The relations between the count ratio of XGRE1 to XGRE3 and of XGRE2 to XGRE3.]
		{The relations between the count ratio of XGRE1 to XGRE3 and of XGRE2 to XGRE3 with various elevation and azimuth angles, when assuming a typical TGF photon spectrum (see the main text).
		The error for the 90$^{\circ}$ elevation angle indicates a statistical one when considering a typical TGF fluence of 0.65~photons~cm$^{-2}$.
		Angles described in the plot are azimuth angles.\label{fig:ang_res}}
		\end{center}
	\end{figure} 

	Figure~\ref{fig:ang_var} shows the change in the number of counts of each sensor with respect to the azimuth angle. 
	The number of counts is normalized by the total number of counts of the three sensors at each elevation angle. 
	The count rate of each sensor changes like a sine curve with respect to the azimuth angle, 
	The maximum count values for XGRE1, 2, and 3 at an elevation angle of 60 $^{\circ}$ are at azimuth angles of 300$^{\circ}$, 60$^{\circ}$, and 170$^{\circ}$, respectively. 
	The minimum values are at 110$^{\circ}$, 250$^{\circ}$, and 0$^{\circ}$, respectively. 
	The amount of changes increases as the elevation angle decreases.

	A practical method for angle detection is analysis on two-dimensional plots.
	Figure~\ref{fig:ang_res} shows the ratio of the counts of XGRE1 and 2 divided by the count of XGRE3 in the two-dimensional plane. 
	On this plot, the ratio makes an oval orbit as the azimuth changes. The smaller the elevation angle, the larger the oval orbit. 
	Since the orbits are not superimposed, the elevation and azimuth angles can be uniquely determined by two numbers, 
	the count ratio of XGRE1 and XGRE2 to XGRE3, if the count data has enough statistics. 
	On the other hand, in practice, the angle estimation accuracy is limited by statistical errors. 
	The error bar attached to the 90$^{\circ}$-elevation result shows a statistical uncertainty at one standard deviation, when considering the typical TGF fluence of 0.65~photons~cm$^{-2}$ (0.04--10~MeV).
	The error bar is superimposed on the orbit of the two-dimensional plane at an elevation angle of 60$^{\circ}$.
	Therefore, the azimuth angle cannot be estimated with the typical-fluence TGFs when the elevation angle is 60$^{\circ}$ or above.
	The estimation accuracy of azimuth angle at an elevation angle of 45$^{\circ}$ is about $\pm$30$^{\circ}$ near azimuth 0$^{\circ}$, and about $\pm$50$^{\circ}$ near azimuth 180$^{\circ}$.
	The estimation accuracy of azimuth angle will be improved for brighter TGFs.

\section{Conclusion}\label{sect:concl}
	XGRE was the key instrument of TARANIS for high-energy atmospheric phenomena such as TGFs and TEBs.
	It consisted of three sensors with LaBr$_{3}$ and plastic scintillators, and was sensitive to X-rays, gamma rays, and electrons. 
	The assembly of the flight model started in February 2018, and validation tests of the sensors were performed in Paris, before its delivery to CNES Toulouse in November 2018.
	XGRE was then mounted on the satellite, and experienced environment and calibration tests with the satellite.

	Based on the thermal-cycle test performed from April to May 2018, the gain variation by the temperature of XGRE was measured.
	The correlation between gain and temperature was clear above 20$^{\circ}$C, but unclear below 20$^{\circ}$C.
	During cold cycles, the energy resolution degraded and sometimes one emission line from the calibration source was split into two peaks
	because the two high-voltage modules in one sensor seemed to have slightly different dependence to temperature.

	The calibration test with XGRE onboard the satellite was performed from February to March 2020 at CNES Toulouse.
	The energy range of XGRE for X-ray and gamma rays was determined to be 0.04--11.6~MeV, 0.08--11.0~MeV, and 0.08--11.3~MeV for XGRE1, 2, and 3, respectively.
	The energy resolution (FWHM) at 0.662~MeV was 20.5\%, 25.9\%, and 28.6\%, respectively.

	To investigate the performances of XGRE, a simulation model was developed and Monte Carlo simulations were performed with Geant4.
	The model was verified by comparing the result of the calibration test with the satellite, and its uncertainty was less than 15\%.
	The effective area of XGRE, derived from a Monte Carlo simulation with the verified model, reached the maximum at 0.1 MeV, 838~cm$^{2}$, 
	and the minimum at 3.5~MeV, 191~cm$^{2}$, in total for the three sensors. 
	The effective area below 0.3~MeV was sensitive to the elevation and azimuth angle of incident X-rays and gamma rays.
	Therefore, the arrival angle of photons can be estimated with the ratio of photon counts measured by the three sensors, 
	although the accuracy was limited by the statistical uncertainty of photon counts.

	TARANIS was launched on 17 November 2020. However, the mission was lost due to the failure of the launcher.
	Although XGRE was unable to directly contribute to high-energy atmospheric physics,
	the knowledge obtained during the development of XGRE will be valuable for a future mission of high-energy atmospheric physics and high-energy astrophysics.


\subsection*{Disclosures}
	The authors declare no conflicts of interest.

\subsection* {Acknowledgments}
	TARANIS was the satellite mission led by CNES. The XGRE team is strongly supported by CNRS (Centre National de la Recherche Scientifique) and CNES.
	Y.W. is supported by MEXT-KAKENHI Grants of Japan, No.18J13355 and 20K22354, 
	RIKEN Special Postdoctoral Research Program, and International Joint Research Program of Institute for Space-Earth Environmental Research, Nagoya University. 

\subsection* {Data, Materials, and Code Availability} 
	The materials to produce figures in this paper are published and can be obtained on Mendeley Data
	(\url{https://doi.org/10.17632/4smf49dyrw.1}). The Geant4 model is available upon request to Y.W. or D.S.

\bibliographystyle{spiejour}


\vspace{2ex}\noindent\textbf{Yuuki Wada} is an assistant professor at the Graduate School of Engineering, Osaka University.
	He received his PhD in physics from the University of Tokyo in 2020. 
	He was a visiting researcher at the Astroparticle and Cosmology (APC) laboratory in Paris, France, in 2018 and 2020, 
	working for on-ground calibration campaigns of the TARANIS/XGRE. 
	His current research interests include high-energy atmospheric physics, meteorology, radar remote sensing, and radiation measurements.
	
\vspace{2ex}\noindent\textbf{Philippe Laurent} is employed at the CEA Astrophysical Department in Saclay and Astroparticle and Cosmology (APC) laboratory in Paris, France. 
	He specializes in high-energy astrophysics and got his PhD in 1992, studying high-energy emission from black holes X-ray binaries. 
	He has been participating in the ESA/INTEGRAL gamma-ray mission since 1995 and is now the INTEGRAL/IBIS telescope Co-PI. 
	He was also a PI of the TARANIS/XGRE instrument developed at the APC laboratory.
	
\vspace{2ex}\noindent\textbf{Ion Cojocari} was an instrumentation engineer at the Astroparticle and Cosmology (APC) laboratory,
	having participated in the integration, testing, and calibration of the XGRE instrument.
	He obtained his PhD in physics, developing Compton polarimeters dedicated to measuring gamma-ray polarization. 
	He is currently employed at IJCLab, developing low-threshold cryogenic germanium detectors 
	to be used in the next generation of low-mass dark matter search experiments.
	
\vspace{2ex}\noindent\textbf{St\'{e}phane Colonges} is an expert in reliability engineering at the APC laboratory of CNRS. 
	He was involved in the TARANIS mission as a product assurance engineer.
	
\vspace{2ex}\noindent\textbf{David Sarria} is a researcher at the Birkeland Centre for Space Science, the University of Bergen in Norway. 
	He received his PhD from the Institut de recherche en astrophysique et planétologie at the University of Toulouse in 2015 and 
	is a post-doctoral researcher at the Astroparticle and Cosmology (APC) laboratory at the University of Paris Diderot on the TARANIS spacecraft,
	particularly on Monte-Carlo modeling of the XGRE instrument.
	
\vspace{2ex}\noindent\textbf{Kazuhiro Nakazawa} is an associate professor at Nagoya University, focusing on X-ray and
	MeV gamma-ray astronomy in both observational and detector development aspects, 
	including on-ground thundercloud gamma-ray observations. He received his PhD from the University of Tokyo in 2001. 
	He has long experience in in-orbit crystal scintillator detector development.

\vspace{1ex}
\noindent Biographies and photographs of the other authors are not available.

\listoffigures
\listoftables

\end{spacing}
\end{document}